\documentclass[twocolumn]{aastex62}
\usepackage{natbib,amssymb,amsmath,graphicx}
\usepackage{textcomp}
\usepackage{longtable}
\usepackage{subfigure}

\shorttitle{Frequent flare on eclipsing binary BX Tri}
\shortauthors{Luo et al.}
\begin{document}

\title{ Frequent flare events on the short-period M-type eclipsing binary BX Tri}
\author{ChangQing Luo}
\affiliation{Key Laboratory of Optical Astronomy, \\
National Astronomical Observatories, Chinese Academy of Sciences, \\Datun Road 20A,
Beijing  100101, China;}
\author{XiaoBin.Zhang}
\affiliation{Key Laboratory of Optical Astronomy, \\
National Astronomical Observatories, Chinese Academy of Sciences, \\Datun Road 20A,
Beijing  100101, China;}
\author{Kun Wang}
\affiliation{Department of Astronomy, China West Normal University, Nanchong 637002, China;}
\author{Chao, Liu}
\affiliation{Key Laboratory of Optical Astronomy, \\
National Astronomical Observatories, Chinese Academy of Sciences, \\Datun Road 20A,
Beijing  100101, China;}
\author{Xiangsong Fang}
\affiliation{Chinese Academy of Sciences South America Center for Astronomy, \\
National Astronomical Observatories, CAS, Beijing 100101, China;}
\affiliation{Instituto de Astronom\'{i}a, Universidad Cat\'{o}lica del Norte, Av. Angamos 0610, Antofagasta, Chile;}

\author{Chunguang Zhang}
\affiliation{Key Laboratory of Optical Astronomy, \\
National Astronomical Observatories, Chinese Academy of Sciences, \\Datun Road 20A,
Beijing  100101, China;}

\author{Licai Deng}
\affiliation{Key Laboratory of Optical Astronomy, \\
National Astronomical Observatories, Chinese Academy of Sciences, \\Datun Road 20A,
Beijing  100101, China;}
\author{ Jundan Nie}
\affiliation{Key Laboratory of Optical Astronomy, \\
National Astronomical Observatories, Chinese Academy of Sciences, \\Datun Road 20A,
Beijing  100101, China;}
\author{ Lester Fox-Machado}
\affiliation{Observatorio Astron\'omico Nacional,  Instituto de Astronom\'{\i}a--Universidad Nacional Aut\'onoma de M\'exico, Ensenada BC, M\'exico;}
\author{ Yangping Luo}
\affiliation{Department of Astronomy, China West Normal University, Nanchong 637002, China;}
\author{ Hubiao Niu}
\affiliation{Department of Astronomy, Beijing Normal University, Beijing, China;}

\begin{abstract}
We present long-term, multi-color photometric monitoring and spectroscopic  observations of the short-period M-type eclipsing binary BX Tri. Six flare events were recorded in 4 years from 2014 to 2017.  Three of them were detected on one night within an orbital cycle.  The strongest one was identified on December 23, 2014. With the amplitudes of $\Delta$B=0.48 mag, $\Delta$V=0.28 mag,  $\Delta$R=0.10 mag, $\Delta$I=0.02 mag, the total energy due to the flare event was measured to be 4.08 ($\pm$0.24) $\times$$10^{34}$ erg, exceeding the superflare level ($10^{34}$). Based on the observations, the evolutionary status of the binary system as well as the long-term orbital period changes were analyzed.  It reveals that BX Tri is probably a semi-detached system with the primary component filling its Roche lobe. The extremely high occurrence of flare events of the binary could be related to the rapid mass transfer between components.

\end{abstract}


\section{INTRODUCTION}
Stellar flares are violent events of sudden energy releases in stellar atmospheres (Benz $\&$ Gudel 2010, Kowalski et al. 2010; Shibayama et al. 2013). These remarkable stellar activities are common on M dwarfs, presumably caused by magnetic reconnection in their atmospheres. Rotational periods and convective envelope depths are also suggested related to this phenomena (Reiners et al. 2012; Browning 2008; Silvestri et al. 2005). Recently, a large number of  M-type stars with flare activities have been examined based on Kepler data (Hawley et al. 2014;  Lurie et al. 2015; Silverberg et al. 2016 and  Davenport et al. 2014). The statistical characteristics of the flares for M dwarfs have presented strong correlations among the flare energy, duration, and amplitude and found that late-type M dwarfs have frequent flares (Yang et al. 2017; Chang et al. 2017;  Hawley et al. 2014, Shibata et al. 2013). 

Flare events have also been detected in close binaries. The strong interaction between the components in a binary system will  typically generate magnetic activity (Noyes et al. 1984). Recently, 234 flare binaries out of the 1049 binaries were identified in the Kepler Eclipsing Binary Catalog (Gao et al. 2016). Seven flares were detected from five RS CVn-type binaries (Pandey $\&$ Singh 2012). A few close binaries with flare events have been detected and the frequency, energy of the flares were studied in detail, such as  EV Lac (Honda et al. 2018),  GSC 02314-00530 (Dimitrov \& Kjurkchieva (2010), GJ 3236 (Smelcer et al. 2017), CU Cnc (Qian et al. 2012) and CM Dra (Nelson \& Caton, 2007).

The present paper concerns long-term flare monitoring of the M-type eclipsing binary BX Tri (also 1SWASP J022050.85+332047.6, GSC 02314-00530).  We chose this target because it was classfited as a W UMa binary with a short period 0.19263 day (Terrell 2014).  The short-period limit of contact binaries has been reported at about 0.22 d (Rucinski 2007), and were widely discussed both observationally and theoretically (Rucinski \& Pribulla 2008, Jiang et al. 2012; Zhang et al, 2014).  Obviously, BX Tri can be a good sample for the study of the limit period of contact binary. This target was also reported with its coincidence with the ROSAT X-ray source 1RXS J022050.7+332049 (Norton et al. 2007). Recently, flare activities have been detected by Dimitrov \& Kjurkchieva  (2010). Combining the radial-velocity curves,  they determined the absolute physical parameters of BX Tri as $M_1= 0.51 M_\odot$; $M_2=0.26 M_\odot$; $R_1=0.55 R_\odot$; $R_2~=0.29 R_\odot$; $L_1~=0.053 L_\odot$; $L_2~=0.007 L_\odot$.  The following linear ephemerides were also derived by those authors:

 HJD(Min I) = 2451352.061633 + 0.1926359 $\times$ E.
 
 The flare activities of BX Tri were observed by Han (2015).  And the magnetic activities of it was discussed by Zhang, Pi \& Yang (2014). Their results revealed that BX Tri has strong magnetic activity. All of the information implies that BX Tri is a very active binary with M-type components and a very short period. Therefore, we carried out long-term, multi-color photometric monitoring and spectroscopic  observations in order to study the flare characteristic of BX Tri. The details of observations and data reduction are given in Section 2. The analysis of photometric solution is performed in Section 3. The behavior of the flare events is draw in Section 4.  Finally,  we give our discussion and conclusion in Section 5.

\section{OBSERVATIONS AND DATA REDUCTION}
\subsection{ New photometry }

The key to the study of flares on active stars is sufficient observations, so that complete samples can be collected. Therefore, we made a detailed monitoring observations for BX Tri from 2014 to 2017 including photometric and spectroscopic observations. The new photometric observations of BX Tri were carried out  in B-, V-, R-, and I- bands with three small telescopes: the 85-cm reflecting telescope at Xinglong Station of NAOC\footnote{National Astronomical Observatories, Chinese Academy of Sciences.} (hereafter XL 85cm),  50BiN prototype telescope at Delingha of PMO (hereafter DLH 50BiN) \footnote{Purple Mountain Observatory, Chinese Academy of Sciences}, and  the 1-m telescope at Nanshan Station in Xingjiang (hereafter XJ 1m).  The details of the telescopes and the data reduction are as follows:

XL 85cm was equipped with  an Andor 2K$\times$2K CCD  camera and a standard  Johnson - Cousins - Bessel multicolor filter system (Zhou et al. 2009). The effective field of view is $35' \times 35' $and  the pixel scale is $0''.96$ pixel$^{-1}$. DLH 50BiN has two parallel optical systems. Each has an Andor 2K$\times$2K CCD camera and provides a $20' \times 20'$ field of view with an pixel scale of about 0$''$.59 pixel$^{-1}$. Two standard Johnson-Cousins-Bessell BV filters were applied to simultaneous two-color photometry (Deng et al. 2013; Tian et al. 2016). XJ 1m  is equipped with a 4K$\times$4K CCD camera gives a 1.3$^{\circ}$ $\times$ 1.3$^{\circ}$ field of view with resalution of about $1''.13 $pixel$^{-1}$ (Liu 2014). 
A total number of 11 observational nights were obtained, 8 nights from XL 85cm with  BVRI bands, 2 nights from DLH 50BiN with BV bands, and 1 night from XJ 1m with BV bands, respectively.   The journal of the photometric observations are list in Table1.

All images were reduced preliminarily with the standard process, including subtracting the bias and dividing flat fields from the object frames.  Then the instrument magnitudes of the stars were extracted from these images using aperture photometry of the IRAF.  A star near the variable star was chosen as the comparison star (TYC 2314-1655), and has a similar brightness and color with the variable star. Another star in the same field of view was selected as the check star (TYC 2314-780). The differential magnitudes of these stars were extracted in each frame. Then, the instrumental magnitudes were converted to the standard system with the standard stars (comparison star,TYC 2314-1655, V = 12.4mag, B - V = 0.22 mag, V - R = 0.20 mag, V - I = 0.46 mag, Dimitrov \& Kjurkchieva (2010) .  Finally, 11 nights data were obtained in total and 6 flares were detected in 4 nights, November 22, 23, 2014 and November 10, 12, 2017, respectively.  Three of them were detected in one night in one period on November 22, 2014, the other three from remaining three days. The light curves with flares are shown in Figure 1, where the yellow arrow indicates the flares places.  \textbf{Focusing on the B-band and V-band light curve, the color of  B - V were calculated, shown below the light curve in Figure 1. The color at the peak of all flares were also presented in Table 6. } We also calculated the epochs of minimum light with mean values of the B-, V-, R-, and I- filters determined by applying the K-W method (Kwee $\&$ Van Woerden 1956),  as given in Table 2. 

\begin{figure*}[t]\setcounter{figure}{0}
 \begin{minipage}[c]{0.52\textwidth}
  \centering
  \includegraphics[width=4.in]{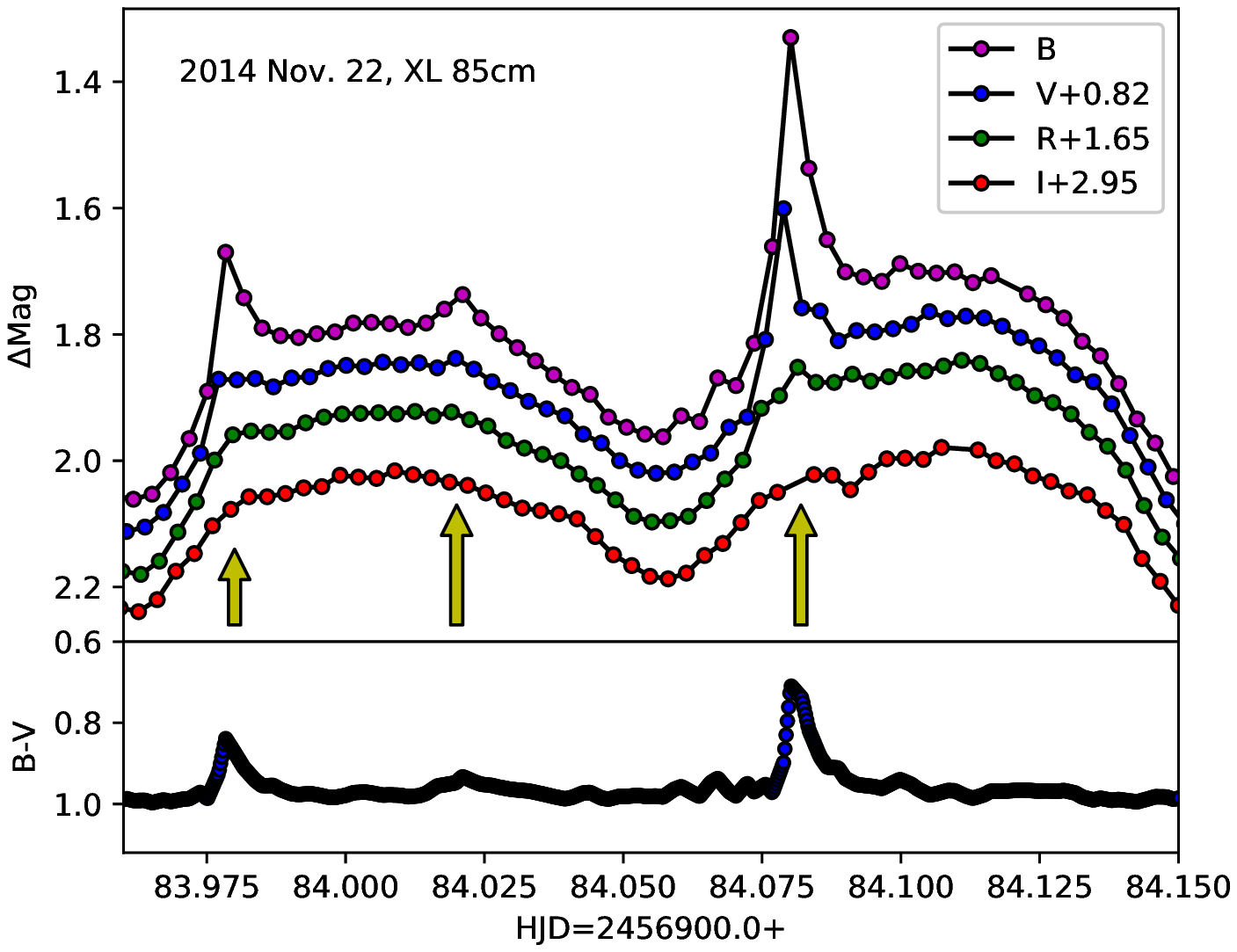}%
  \hspace{0.1in}%
 \end{minipage}%
 \begin{minipage}[c]{0.52\textwidth}
  \centering
  \includegraphics[width=4.in]{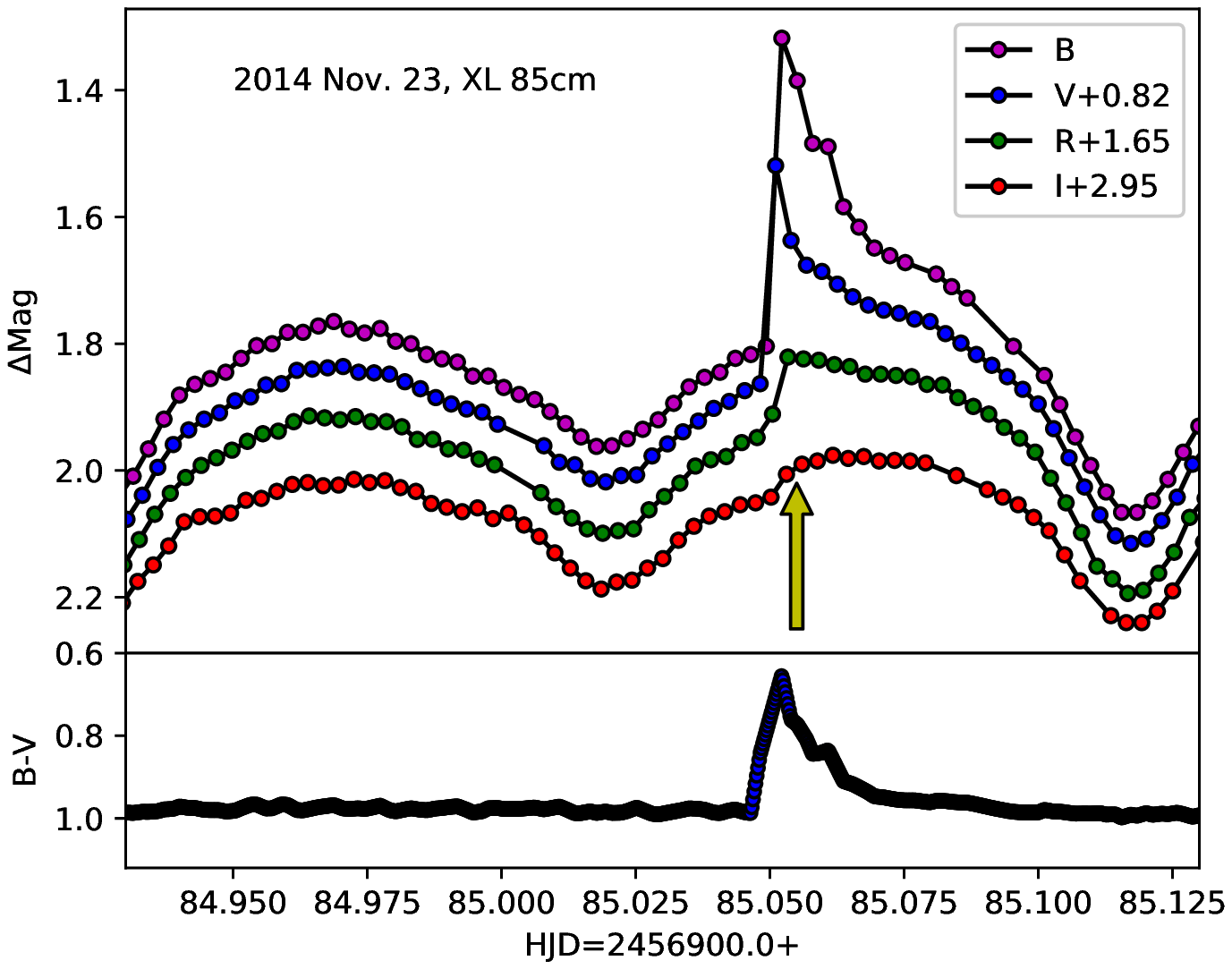}%
  \hspace{1in}%
 \end{minipage}
\begin{minipage}[c]{0.52\textwidth}
  \includegraphics[width=4.in]{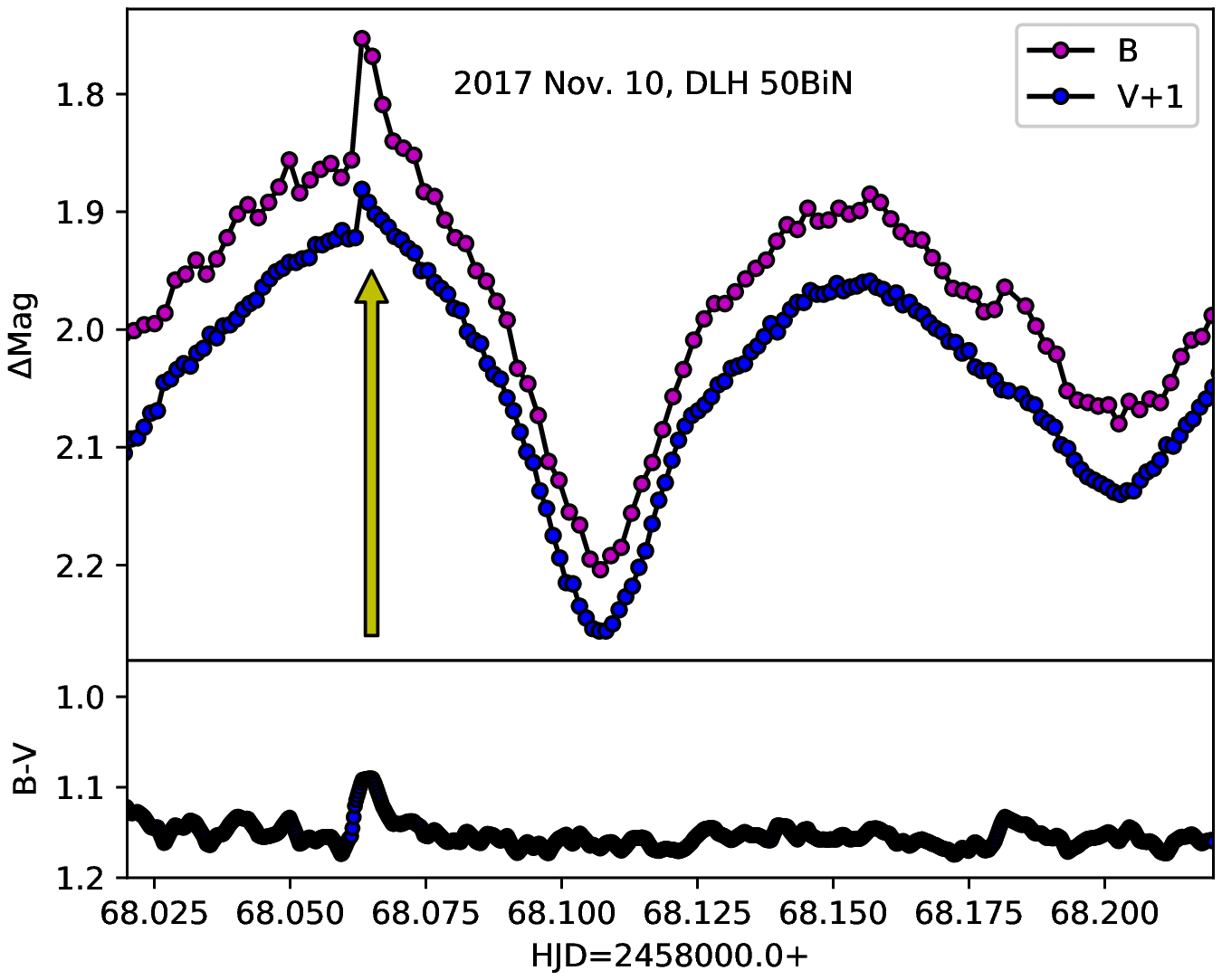}%
  \hspace{1in}%
 \end{minipage}%
 \begin{minipage}[c]{0.52\textwidth}
  \includegraphics[width=4.in]{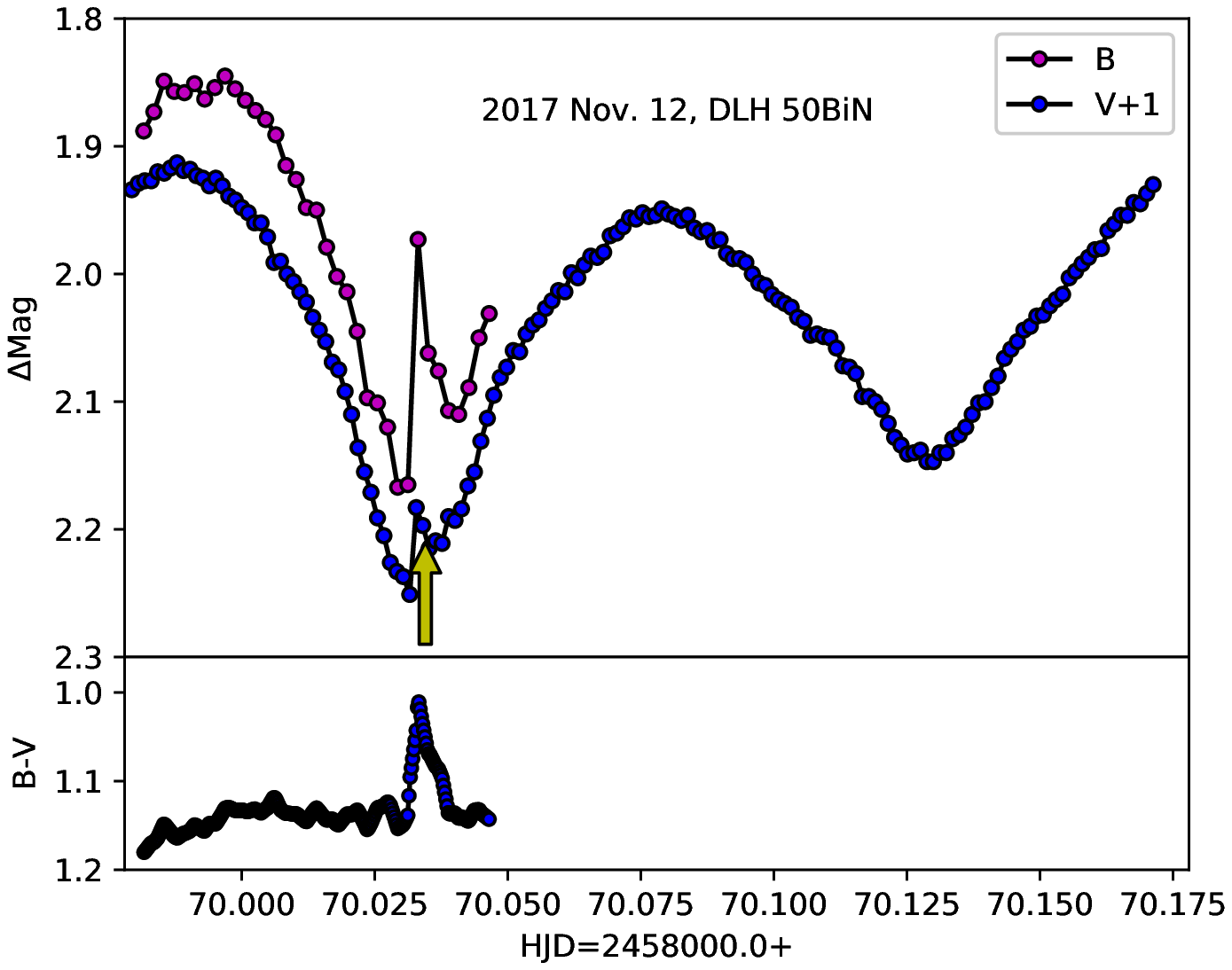}%
  \hspace{0.1in}%
 \end{minipage}%
 \caption{Photometric observations of BX Tri on November 22 (top-left) and 23 (top-right), 2014 with flares in B, V, R, I bands and  November 10 (bottom-left) and 12 (bottom-right), 2017 with flares in B-, V- bands. The yellow arrow indicates the flares places.}
\end{figure*}

\begin{table*}
\begin{center}
\caption{Journal of the Time-series CCD Photometric for BX Tri.}\label{tab1}
\def\temptablewidth{0.67\textwidth}
{\rule{\temptablewidth}{0.5pt}}
\begin{tabular*}{\temptablewidth}{@{\extracolsep{\fill}}  lccrrrc}
\hline
Date (UT)        & HJD (start)    & HJD (end)  &   Filter   &  Exposure(s)& $N_{obs}$     & Telescope\\
                       & 2450000.00+              &2450000.00+              &               &                 &     & \\
\hline                                               
\cline{1-2}
2014 Nov. 22   &6983.91 &6984.25              &B/V/R/I        &  90/60/20/15     & 102         &  XL 85cm     \\  
2014 Nov. 23   &6984.92  &6985.15             & B/V/R/I       &100/65/20/15     & 90           &  XL 85cm     \\  
2016 Oct. 10   &7672.02  &7672.22               &B/V/R/I       &90/60/20/15       &  83           &  XL 85cm     \\ 
2016 Nov. 22   &7714.95  &7715.35             & B/V/R/I      &90/60/20/15        &  144          & XL 85cm     \\ 
2016 Nov. 23   &7716.06   &7716.41            &B/V/R/I       &90/60/20/15        & 166          &  XL 85cm     \\
2016 Dec. 14   & 7736.93  &7737.22           & B/V/R/I      &150/80/25/20      & 101          &  XL 85cm     \\
2016 Dec. 15   &7737.91   &7738.22          & B/V/R/I      &90/60/20/15       & 108          & XL 85cm     \\

2017 Nov. 10   &8068.01    &8068.35         &  B/V           &160/100             &179           & DLH 50BiN       \\
2017 Nov. 12   &8069.97   &8070.04           &  B/V           & 160/100            & 36            & DLH 50BiN       \\
2017 Nov. 13   &8071.17   &8071.22           &  B/V           &90/60                 & 71            & XL 85cm     \\
2017 Nov. 13   &8071.25    &8071.43           & B/V            &80/35        & 84            & XJ 1m     \\

\end{tabular*}
{\rule{\temptablewidth}{0.5pt}}                                                                                                  
\begin{minipage}[t]{0.67\textwidth}                                                                                              
\end{minipage}                                                                                                                   
\end{center}  
\end{table*}


\begin{table}
  \caption{New times of light minimum for XZ Leo observed in this work.}
   \begin{center}
   \begin{tabular}{lccc}\hline
HJD &Error     &Type &Filter \\
\hline\noalign{\smallskip}
2456983.96063 & 0.0001 & I & $BVRI$\\
2456984.05591 & 0.0001 & II & $BVRI$\\
2456984.15436 & 0.0001 & I & $BVRI$\\
2456984.92548 & 0.0001 & I &$BVRI$\\
2456985.01941 & 0.0002 & II & $BVRI$\\
2456985.11725 & 0.0001 & I &$BVRI$\\
2457672.05090& 0.0001 & I &$BVRI$\\
2457672.14557& 0.0002 & II & $BVRI$\\
2457672.23822& 0.0001 & I &$BVRI$\\
2457672.33325& 0.0002 & II & $BVRI$\\

2457736.96613& 0.0001 & I &$BVRI$\\
2457737.06244& 0.0001 & II & $BVRI$\\
2457737.16064& 0.0001 & I &$BVRI$\\

2457737.93018& 0.0001 & I &$BVRI$\\
2457738.02356& 0.0001 & II & $BVRI$\\
2457738.12427& 0.0001 & I &$BVRI$\\
2457738.21786 & 0.0001 & II & $BVRI$\\

2457715.00761& 0.0001 & I &$BVRI$\\
2457715.10107 & 0.0001 & II & $BVRI$\\
2457715.19928& 0.0001 & I &$BVRI$\\
2457715.29358 & 0.0001 & II & $BVRI$\\

2457716.06549& 0.0001  & II & $BVRI$\\
2457716.16132& 0.0001 & I &$BVRI$\\
2457716.25726 & 0.0001 & II & $BVRI$\\
2457716.35638& 0.0001 & I &$BVRI$\\

2458068.10714& 0.0001 & I &$BV$\\
2458068.20259 & 0.0001 & II & $BV$\\
2458068.29805& 0.0001 & I &$BV$\\

2458070.03380& 0.0001 & I &$BV$\\
2458070.12875 & 0.0001 & II & $BV$\\

2458071.18900& 0.0001 & I &$BV$\\

2458071.28200& 0.0001 & I &$BV$\\
2458071.38100 & 0.0001 & II & $BV$\\
\noalign{\smallskip}\hline
  \end{tabular}
  \end{center}
\end{table}

\subsection{ Spectroscopy}

 The spectroscopic observations were carried out with three instruments. The first one is Beijing Faint Object Spectrograph and Camera (BFOSC; Fan et al. 2016)  mounted on 2.16 m telescope at Xinglong station of the National Astronomical Observatories of China (NAOC). \textbf{BFOSC has a CDD with size of 2048 $\times$ 2048. We adopted the  grating E9 + G10 with wavelength range from 3740 \AA ~to 10200 \AA  $~$ and 1$''$.6 width slit. The exposure time of each image was 30 minutes.  The second one is optical median resolution (OMR ) spectrograph, which also mounted on  2.16 m telescope at Xinglong station of  NAOC. We chose the grating of 600lp/mm with the wavelength ranging from 3200 \AA~to 7500 \AA and a slit width of 2$''$.0. The exposure time was 20 minutes. The last one was on the 2.12m telescope at the Observatory Astronomical National on the Sierra San Pedro Mártir (OAN-SPM) . We used a 2048$\times$2048 E2V CCD to collect the high-resolution (the maximum resolution is R = 18,000 at 5000 \AA) ECHELLE spectra at the slit size 1$''$.0. The wavelength range is from 3800 \AA~to 7100 \AA.  The exposure time was 40 minutes. No filter was used during the observations.  The journals of the spectroscopic observations is presented in Table 3. At last, totally of 12 spectra of BX Tri during 2016 December - 2017 November were obtained, which covered the whole orbital phase of BX Tri.}
 
Reduction of the spectra was performed as the standard process by using IRAF/IMRED packages including bias, flat calibration, and cosmic-ray removing. The wavelength calibration was made by taking the spectra of Fe - Ar lamp, He - Ar lamp and Th - Ar lamp for BFOSC, OMR of XL, and ECHELLE of OAN-SPM, respectively. Then, the data reduction was primarily performed using the IRAF/APALL package to obtain the normalized spectrum. Figure 2 presents H$\alpha$, H$\beta$ and H$\gamma$ profiles at different phases. \textbf{We can see that there is a clearly asymmetric in the H alpha line at phase = 0.1854. This spectra was observed by OAN-SPM in Mexico with a  high-resolution. The asymmetric profile is indeed a superposition of emission lines from both stars. It is consistent with the orbital velocity at the phase.}

\begin{table*}
\begin{center}
\caption{Journal of the spectral observations for BX Tri.}\label{tab1}
\def\temptablewidth{0.67\textwidth}
{\rule{\temptablewidth}{0.5pt}}
\begin{tabular*}{\temptablewidth}{@{\extracolsep{\fill}}  lcrrrc}
\hline
\cline{1-2}
Date (UT)        & HJD (start)               & phase         &    Exp(s)      &   Observatory          &Telescope \\
     & 2450000+00            &       &         &             & \\
\hline   
2017 Feb. 09  & 7793.64        & 0.1854        &   2400         & SPM, Mexico           &ECHELL, 2.12m \\
2016 Dec. 29  & 7751.93       & 0.6614         &   1800         &XL, China           &BFOSC, 2.16m \\
2016 Dec. 29  & 7751.95      & 0.7527         &   1800         &XL, China            &BFOSC, 2.16m \\
2017 Nov. 13  & 8071.18       & 0.8980         &   1800         &XL, China          &BFOSC, 2.16m \\
2017 Nov. 13  & 8071.20       & 0.0127         &   1800         &XL, China            &BFOSC, 2.16m \\
2017 Dec. 22  & 8080.03       &0.8731        &   1200         &XL, China           &OMR,  2.16m \\
2017 Dec. 22  & 8080.05       &0.9455        &   1200         &XL, China           &OMR,  2.16m \\
2017 Dec. 22  & 8080.06       &0.0284        &   1200         &XL, China            &OMR,  2.16m \\
2017 Dec. 22  & 8080.08       &0.0903       &   1200         &XL, China            &OMR,  2.16m \\
2017 Dec. 22  & 8080.09       &0.1628        &   1200         &XL, China           &OMR, 2.16m \\
2017 Dec. 23  & 8081.14       &0.5944        &   1200         &XL, China            &OMR,  2.16m \\
2017 Dec. 23  & 8081.15       &0.6692       &   1200         &XL, China            &OMR, 2.16m \\
\end{tabular*}
{\rule{\temptablewidth}{0.5pt}}                                                                                                  

\begin{minipage}[t]{0.67\textwidth}                                                                                              
\end{minipage}                                                                                                                   
\end{center}  
\end{table*}

\begin{figure*}
\center
\includegraphics[scale=.5]{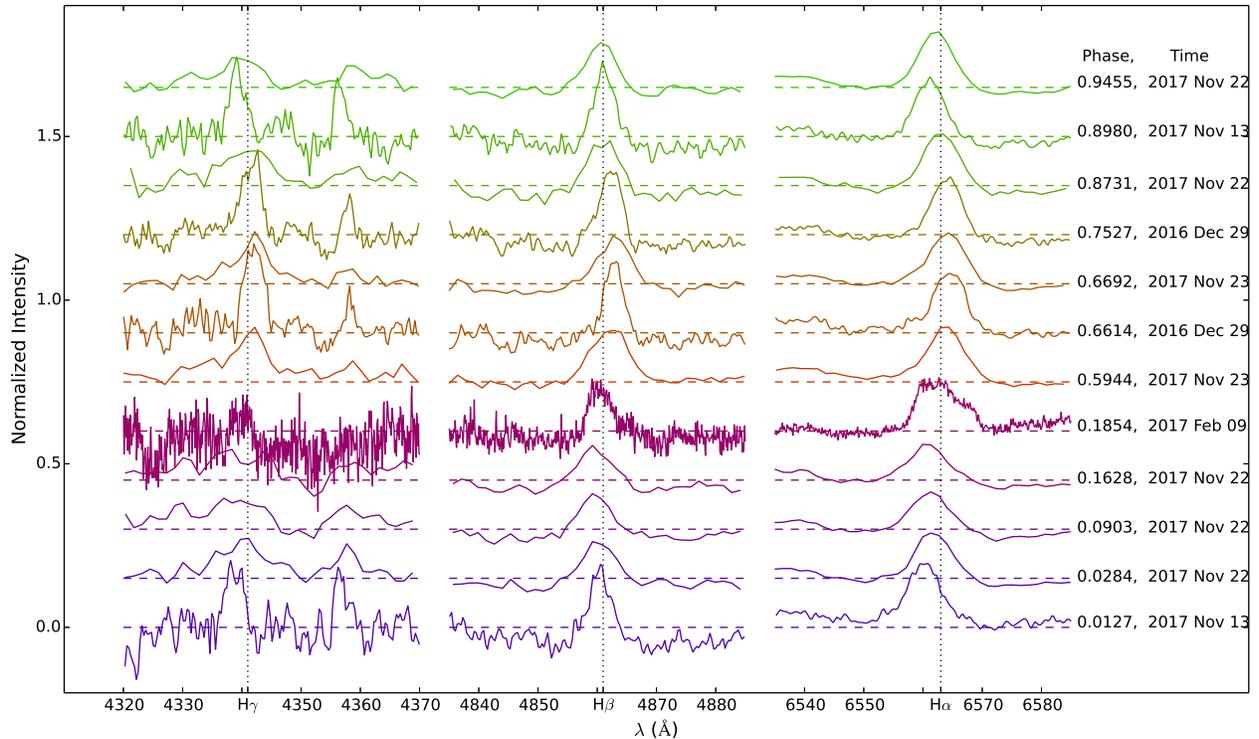}
\center
\caption{The totally of 12 spectra of BX Tri during 2016 December - 2017 November are presented, which covered the whole orbital phase of BX Tri. The three dash line show the H$\alpha $, H$\beta$ and H$\gamma$ profiles at the different phases. }
\center
\label{fdiff}
\end{figure*}

\section{ analysis of the photometric data}

\subsection{Period study}
We have collected  all of the available times of light minimum from database O-C gateway (http://var.astro.cz/ocgate/) and the literatures to calculate the orbital period. Combining 33 times of light minima observed by us, 131 times of light minima are collected in total. All of the  data are observed using CCD method. Table 4 presents all the times of light minima and  the types of eclipses, where "p" and "s" refer to the primary minima and the secondary minima respectively. The corresponding circles Epoch and O-C values are based on the new linear ephemeris:
 \begin{eqnarray}
\begin{aligned}
Min.I =&2457737.93018(4) + 0^d.1926359 \times E.\\ 
\end{aligned}
\end{eqnarray}
 
Finally, together with all of the eclipse time data, the following quadratic ephemeris was derived by using the least-square method:
\begin{eqnarray}
\begin{aligned}
Min.I =&2457737.93003(4) + 0^d.1926368(5) \times E\\ 
&-3.76 \times 10^{-11}  \times E^2.
\end{aligned}
\end{eqnarray}

 The values of epoch and O-C are showing in Table 4. The O-C diagram for BX Tri is plotted  in the upper panel of Figure 3 with solid lines. The observed data are plotted with open circles. The bottom panel displays the residuals between the ephemeris and observed data. It seems that there is a continuous secular decreasing in this system. The period decreasing rate $dP/dt=-1.42 \times 10^{-7}$  $ days~ yr^{-1}$ was derived from equation (2). Since it is composed of two late-type components, BX Tri belongs to chromospheric active binaries (CAB).  This type of systems usually have strong magnetic activities. The period decreasing may be due to the lose of angular momentum via a magnetized wind, also called magnetic braking (Stepien et al. 1995; Demircan et al. 1999; Luo et al. 2010). The mass transfer can also lead to the period decreasing.  \textbf{By using WD model simulating, BX Tri system was probably a Semi-detached binary with primary component filling its Roche lobe (see in section 3.2). Therefore, mass transfer from primary to secondary components can lead to period decreasing in a long timescales. So the period decreasing of BX Tri is most likely cause by the mass transfer from primary to secondary  component.}

\begin{table*}
  \caption{CCD Times of light minimum for BX Tri. }
   \begin{center}
   \fontsize{7pt}{7pt}
   \selectfont
   \begin{tabular}{lccccllccccl}\hline
No.&HJD             &Type     &Epoch      & O-C       & References                &$~~~~~~~ $NO.& HJD             &Type &Epoch   & O-C       & References            \\
&2400000+.   &             &                &               &                                   &$~~~~~~~ $&2400000+.    &         &              &               &                                \\
\hline\noalign{\smallskip}
1& 55126.4750  & s	  &-13556.5   & -0.0036   &   Dimitrov \& Kjurkchieva (2010)  &$~~~~~~~ $67&56719.2830  & p	 &-5288.0   &  0.0041   &   O-C gateway$^{*}$   \\
2& 55126.5700  & p	  &-13556.0   & -0.0049   &   Dimitrov \& Kjurkchieva (2010)  &$~~~~~~~ $68 &56721.3023  & s	 &-5277.5   &  0.0007   &   O-C gateway$^{*}$   \\
3& 55149.2040  & s	  &-13438.5   & -0.0055   &   Dimitrov \& Kjurkchieva (2010)  &$~~~~~~~ $69&56721.3044  & s	 &-5277.5   &  0.0028   &   O-C gateway$^{*}$   \\      
4& 55149.2990  & p	  &-13438.0   & -0.0068   &   Dimitrov \& Kjurkchieva (2010) &$~~~~~~~ $70 &56726.3117  & s	 &-5251.5   &  0.0016   &   O-C gateway$^{*}$   \\     
5& 55149.3990  & s	  &-13437.5   & -0.0031   &   Dimitrov \& Kjurkchieva (2010) &$~~~~~~~ $71 &56726.3128  & s	 &-5251.5   &  0.0027   &   O-C gateway$^{*}$    \\
6& 55149.4930  & p	  &-13437.0   & -0.0055   &   Dimitrov \& Kjurkchieva (2010) &$~~~~~~~ $72 &56983.9606  & p	 &-3914.0   &  0.0015   &   this paper    \\
7& 55156.4270  & p	  &-13401.0   & -0.0063   &  Dimitrov \& Kjurkchieva (2010) &$~~~~~~~ $73 &56984.0559  & s	 &-3913.5   &  0.0005   &this paper   \\    
8& 55156.5260  & s	  &-13400.5   & -0.0036   &   Dimitrov \& Kjurkchieva (2010) &$~~~~~~~ $74 &56984.1543  & p	 &-3913.0   &  0.0026   &   this paper     \\
9& 55525.0383  & s	  &-11487.5   & -0.0016   &   O-C gateway$^{*}$  & $~~~~~~~ $75&56984.9254  & p	 &-3909.0   &  0.0032   &   this paper   \\
10& 55525.1327  & p	  &-11487.0   & -0.0035   &   O-C gateway$^{*}$  &$~~~~~~~ $76&56985.0194  & s	 &-3908.5   &  0.0008   &   this paper     \\
11& 55525.2308  & s	  &-11486.5   & -0.0017   &   O-C gateway$^{*}$  &$~~~~~~~ $77 &56985.1172  & p	 &-3908.0   &  0.0023   &   this paper    \\
12& 55589.9563  & s	  &-11150.5   & -0.0015   &   O-C gateway$^{*}$	 &$~~~~~~~ $78&57050.3181  & s	 &-3569.5   & -0.0036   &   O-C gateway$^{*}$    \\
13& 55590.0512  & p	  &-11150.0   & -0.0029   &   O-C gateway$^{*}$	 &$~~~~~~~ $79 &57050.3192  & s	 &-3569.5   & -0.0025   &   O-C gateway$^{*}$    \\
14& 55594.0022  & s	  &-11129.5   & -0.0009   &   O-C gateway$^{*}$	 &$~~~~~~~ $80 &57241.5139  & p	 &-2577.0   &  0.0022   &   O-C gateway$^{*}$     \\
15& 55594.0964  & p	  &-11129.0   & -0.0030   &   O-C gateway$^{*}$	 &$~~~~~~~ $81&57278.4989  & p	 &-2385.0   &  0.0013   &   O-C gateway$^{*}$     \\
16& 55835.4703  & p	  & -9876.0   & -0.0005   &   O-C gateway$^{*}$	 &$~~~~~~~ $82&57294.3870  & s	 &-2302.5   & -0.0030   &   O-C gateway$^{*}$     \\
17& 55835.4710  & p	  & -9876.0   &  0.0002   &   O-C gateway$^{*}$	 &$~~~~~~~ $83&57294.4872  & p	 &-2302.0   &  0.0009   &   O-C gateway$^{*}$   \\
18& 55838.4592  & s	  & -9860.5   &  0.0026   &   O-C gateway$^{*}$	 &$~~~~~~~ $84&57301.4226  & p	 &-2266.0   &  0.0015   &   O-C gateway$^{*}$   \\
19& 55839.4205  & s	  & -9855.5   &  0.0007   &   O-C gateway$^{*}$	 &$~~~~~~~ $85&57301.5153  & s	 &-2265.5   & -0.0021   &   O-C gateway$^{*}$   \\  
20& 55839.5166  & p	  & -9855.0   &  0.0005   &   O-C gateway$^{*}$	 &$~~~~~~~ $86 &57301.5163  & s	 &-2265.5   & -0.0011   &   O-C gateway$^{*}$   \\  
21& 55839.6150  & s	  &-9854.5   &  0.0026   &   O-C gateway$^{*}$ 	&$~~~~~~~ $87 &57310.4761  & p	 &-2219.0   &  0.0011   &   O-C gateway$^{*}$    \\
22& 55866.4883  & p	  &-9715.0   &  0.0033   &   O-C gateway$^{*}$ 	&$~~~~~~~ $88  &57319.5286  & p	 &-2172.0   & -0.0002   &   O-C gateway$^{*}$    \\
23& 55866.5833  & s	  &-9714.5   &  0.0020   &   O-C gateway$^{*}$ 	&$~~~~~~~ $89 &57319.5290  & p	 &-2172.0   &  0.0002   &   O-C gateway$^{*}$    \\ 
24& 55872.4552  & p	  &-9684.0   & -0.0014   &   O-C gateway$^{*}$ 	&$~~~~~~~ $90  &57319.5292  & p	 &-2172.0   &  0.0004   &   O-C gateway$^{*}$     \\
25& 55872.5542  & s	  &-9683.5   &  0.0012   &   O-C gateway$^{*}$ 	&$~~~~~~~ $91&57328.2913  & s	 &-2126.5   & -0.0024   &   O-C gateway$^{*}$    \\
26& 55905.0106  & p	  &-9515.0   & -0.0013   &   O-C gateway$^{*}$&$~~~~~~~ $92   &57328.3905  & p	 &-2126.0   &  0.0005   &   O-C gateway$^{*}$     \\
27& 55905.1094  & s	  &-9514.5   &  0.0012   &   O-C gateway$^{*}$ 	&$~~~~~~~ $93&57328.4839  & s	 &-2125.5   & -0.0024   &   O-C gateway$^{*}$    \\
28& 55970.3143  & p	  &-9176.0   & -0.0008   &   O-C gateway$^{*}$ 	&$~~~~~~~ $94  &57369.3198  & s	 &-1913.5   & -0.0051   &   O-C gateway$^{*}$    \\
29& 56152.4529  & s	  &-8230.5   &  0.0017   &   O-C gateway$^{*}$ 	&$~~~~~~~ $95&57369.4212  & p	 &-1913.0   &  0.0000   &   O-C gateway$^{*}$    \\
30& 56152.5492  & p	  &-8230.0   &  0.0017   &   O-C gateway$^{*}$ 	&$~~~~~~~ $96  &57387.2390  & s	 &-1820.5   & -0.0009   &   O-C gateway$^{*}$     \\
31& 56515.4742  & p	  & -6346.0   &  0.0028   &   O-C gateway$^{*}$	 &$~~~~~~~ $97 &57387.3368  & p	 &-1820.0   &  0.0006   &   O-C gateway$^{*}$     \\
32& 56557.7566  & s	  & -6126.5   &  0.0019   &   O-C gateway$^{*}$	 &$~~~~~~~ $98&57608.4804  & p	 & -672.0   & -0.0005   &   O-C gateway$^{*}$     \\
33& 56557.7569  & s	  & -6126.5   &  0.0022   &   O-C gateway$^{*}$	 &$~~~~~~~ $99 &57625.4325  & p	 & -584.0   & -0.0003   &   O-C gateway$^{*}$   \\
34& 56557.7577  & s	  & -6126.5   &  0.0030   &   O-C gateway$^{*}$	 &$~~~~~~~ $100 &57625.5269  & s	 & -583.5   & -0.0022   &   O-C gateway$^{*}$   \\
35& 56643.3847  & p	  & -5682.0   &  0.0038   &   O-C gateway$^{*}$	 &$~~~~~~~ $101&57661.4557  & p	 & -397.0   &  0.0003   &   O-C gateway$^{*}$   \\  
36& 56643.3850  & p	  & -5682.0   &  0.0041   &   O-C gateway$^{*}$	 &$~~~~~~~ $102&57661.5497  & s	 & -396.5   & -0.0021   &   O-C gateway$^{*}$   \\  
37& 56643.4775  & s	  & -5681.5   &  0.0003   &   O-C gateway$^{*}$	 &$~~~~~~~ $103&57672.0509  & p	 & -342.0   &  0.0005   &   this paper\\
38& 56643.4798  & s	  & -5681.5   &  0.0026   &   O-C gateway$^{*}$	 &$~~~~~~~ $104&57672.1455  & s	 & -341.5   & -0.0012   &   this paper   \\
39& 56684.2232  & p	  & -5470.0   &  0.0038   &   O-C gateway$^{*}$	 &$~~~~~~~ $105  &57672.2382  & p	 & -341.0   & -0.0048   &   this paper    \\ 
40& 56684.2235  & p	  & -5470.0   &  0.0041   &   O-C gateway$^{*}$	 &$~~~~~~~$106 &57672.3332  & s	 & -340.5   & -0.0061   &   this paper    \\
41& 56684.3175  & s	  & -5469.5   &  0.0018   &   O-C gateway$^{*}$	 &$~~~~~~~ $107&57692.2757  & p	 & -237.0   & -0.0013   &   O-C gateway$^{*}$    \\
42& 56684.3187  & s	  & -5469.5   &  0.0030   &   O-C gateway$^{*}$	 &$~~~~~~~ $108 &57692.3709  & s	 & -236.5   & -0.0024   &   O-C gateway$^{*}$     \\
43& 56684.4144  & p	  & -5469.0   &  0.0023   &   O-C gateway$^{*}$	 &$~~~~~~~ $109 &57715.0076  & p	 & -119.0   & -0.0003   &  this paper   \\
44& 56684.4151  & p	  & -5469.0   &  0.0030   &   O-C gateway$^{*}$	 &$~~~~~~~ $110&57715.1010  & s	 & -118.5   & -0.0032   &   this paper    \\
45& 56692.3125  & p	  & -5428.0   &  0.0024   &   O-C gateway$^{*}$	 &$~~~~~~~ $111 &57715.1992  & p	 & -118.0   & -0.0013   &   this paper    \\
46& 56692.3132  & p	  & -5428.0   &  0.0031   &   O-C gateway$^{*}$	 &$~~~~~~~ $112 &57715.2935  & s	 & -117.5   & -0.0034   &   this paper    \\
47& 56693.2769  & p	  & -5423.0   &  0.0036   &   O-C gateway$^{*}$	 &$~~~~~~~ $113 &57716.0654  & s	 & -113.5   & -0.0020   &   this paper    \\
48& 56693.2779  & p	  & -5423.0   &  0.0046   &   O-C gateway$^{*}$	 &$~~~~~~~ $114 &57716.1613  & p	 & -113.0   & -0.0024   &   this paper     \\
49& 56693.3701  & s	  & -5422.5   &  0.0005   &   O-C gateway$^{*}$	 &$~~~~~~~ $115&57716.2572  & s	 & -112.5   & -0.0028   &   this paper   \\
50& 56693.3702  & s	  & -5422.5   &  0.0006   &   O-C gateway$^{*}$	 &$~~~~~~~ $116&57716.3563  & p	 & -112.0   & -0.0000   &   this paper   \\
51& 56702.2325  & s	  &-5376.5   &  0.0017   &   O-C gateway$^{*}$ 	&$~~~~~~~ $117&57736.9661  & p	 &   -5.0   & -0.0022   &   this paper  \\  
52& 56702.2330  & s	  &-5376.5   &  0.0022   &   O-C gateway$^{*}$ 	& $~~~~~~~ $118&57737.0624  & s	 &   -4.5   & -0.0022   &   this paper   \\  
53& 56702.3295  & p	  &-5376.0   &  0.0024   &   O-C gateway$^{*}$ 	&$~~~~~~~ $119&57737.1606  & p	 &   -4.0   & -0.0003   &   this paper    \\
54& 56702.3298  & p	  &-5376.0   &  0.0027   &   O-C gateway$^{*}$ 	&$~~~~~~~ $120&57737.9301  & p	 &   -0.0   & -0.0013   &   this paper\\
55& 56703.2931  & p	  &-5371.0   &  0.0028   &   O-C gateway$^{*}$ 	&$~~~~~~~ $121&57738.0235  & s	 &    0.5   & -0.0043   & this paper   \\ 
56& 56703.2936  & p	  &-5371.0   &  0.0033   &   O-C gateway$^{*}$ 	&$~~~~~~~ $122&57738.1242  & p	 &    1.0   &  0.0001   &this paper     \\
57& 56709.2642  & p	  &-5340.0   &  0.0023   &   O-C gateway$^{*}$ 	&$~~~~~~~ $123&57738.2178  & s	 &    1.5   & -0.0026   &   this paper   \\
58& 56709.2645  & p	  &-5340.0   &  0.0026   &   O-C gateway$^{*}$ 	&$~~~~~~~ $124&58068.1071  & p	 & 1714.0   & -0.0003   &   this paper     \\
59& 56709.3572  & s	  &-5339.5   & -0.0011   &   O-C gateway$^{*}$ 	&$~~~~~~~ $125&58068.2025  & s	 & 1714.5   & -0.0012   &   this paper    \\
60& 56709.3588  & s	  &-5339.5   &  0.0005   &   O-C gateway$^{*}$ 	&$~~~~~~~ $126&58068.2980  & p	 & 1715.0   & -0.0020   &   this paper    \\
61& 56712.2486  & s	  &-5324.5   &  0.0008   &   O-C gateway$^{*}$ 	&$~~~~~~~ $127&58070.0338  & p	 & 1724.0   &  0.0001   &   this paper \\
62& 56712.3463  & p	  &-5324.0   &  0.0022   &   O-C gateway$^{*}$ 	&$~~~~~~~ $128&58070.1287  & s	 & 1724.5   & -0.0013   &   this paper \\
63& 56714.2737  & p	  &-5314.0   &  0.0033   &   O-C gateway$^{*}$ 	&$~~~~~~~ $129 &58071.1900  & p	 & 1730.0   &  0.0005   &.this paper        \\
64& 56718.3194  & p	  &-5293.0   &  0.0036   &   O-C gateway$^{*}$ 	&$~~~~~~~ $130 &58071.2820  & s	 & 1730.5   & -0.0039   &this paper         \\
65& 56718.3194  & p	  &-5293.0   &  0.0036   &   O-C gateway$^{*}$ 	&$~~~~~~~ $131&58071.3810  & p	 &1731.0    & -0.0012  &this paper \\			
66& 56719.2822  & p	  &-5288.0   &  0.0033   &   O-C gateway$^{*}$ 	& $~~~~~~~ $  &	      	       & 	 &.              &.             &                                      \\
\noalign{\smallskip}\hline
   \end{tabular}
  \end{center}   Note. O-C gateway $^{*}$: http://var.astro.cz/ocgate/.
\end{table*}

\begin{figure}
\center
\includegraphics[scale=.4]{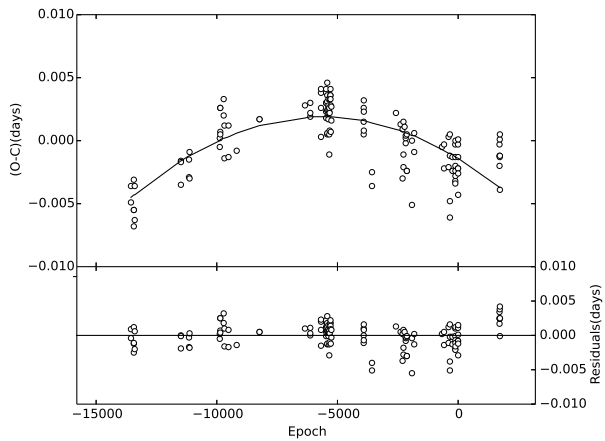}
\center
\caption{O-C curve of BX Tri with all of the observed data (open circales). Upper panel: O-C
diagram computed with Equation (2) with the solid line, indicating that there is a long-term period decrease. 
Bottom panel: resulting residuals.}. 
\center
\label{fdiff}
\end{figure}

\subsection{ Photometric solution of BX Tri}
In order to find out the proper photometric solution, the light curves with and without flare phenomenon are calculated, respectively. Here, our  four-color light curves were simultaneously analyzed by using the Wilson - Devinney (WD) program (Wilson $\&$ Devinney 1971; Wilson 1979, 1990, 1994, 2012; Wilson $\&$ Van Hamme
2003). During the process, we assumed the effective temperature to be T$_1$=3735 K for the primary component, which was determined from the color index and taking into account that the temperature of the primary component T$_1$ is higher than the mean temperature of the binary (Dimitrov \& Kjurkchieva, 2010).  
The initial mass ratio $q$ is fixed to the spectroscopic mass ratio $q$ = 0.509($\pm0.02$) obtained by Rucinski \& Lu (1999).  
The gravity-darkening exponents were set to 0.32 according to the stellar temperatures given by Claret (2000).
The bolometric albedos A1 = A2 = 0.5 (Rucinski 1969) were used because the BX Tri is cool. 
A nonlinear limb-darkening law with a logarithmic form was applied in the light-curves synthesis.
The initial bolometric ($X_1$, $X_2$, $Y_1$, $Y_2$) and monochromatic limb-darkening coefficients ($x_1$, $x_2$, $y_1$, $y_2$,) of the components were taken from Van Hamme (1993).  The adjustable parameters of photometric solution are listed in Table 5. 

We started with light curves without flare activities on Dec 14 and 15, 2016 (named as Solution A without flares). \textbf{In the light curve,  the  light maxima obviously not equal are implying an O'Connell effect , which has been widely suggested that this effect is associated with magnetic activity (O'Connell 1951; Wilsey \& Beaky 2009). Consequently, a cool spot model was introduced  to the primary component.}  At first, mode 2  (detached binary)  was used and  the surface potential of the primary component reached  its Roche limits. Then, we changed to mode 4 (semi-detached)  in the program and obtained the best fit model for the B-, V-, R-, and I- band photometric data simultaneously. The observed data (red, green, blue, purple circles mark B-, V-, R- and I- bands respectively) and the best-fit light curves (black solid lines) are shown in Figure 4. The results and the parameters of the cool spot are shown as solution A in Table 5.
  
\begin{figure}
\center
\includegraphics[scale=.6]{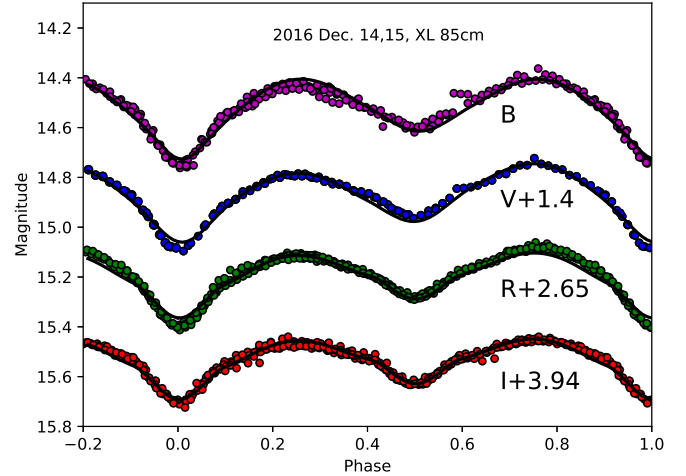}
\caption{Light curves of BX Tri without flares  in the B-, V-, R-, and I- bands obtained using the
XL 85cm on 2016 December 14 and 15. 
 Open circles denote the observational data and the solid line represents the theoretical light curves calculated using with W-D method.}
\center
\label{fdiff}
\end{figure}
  

\begin{table*}
  \caption{The results of photometric solution A, B and C of BX Tri . }
   \begin{center}
   \begin{tabular}{lllll}\hline
   Parameters & 2016 NOV 14,15&2014 DEC 22,23&2017 DEC 10,12&  \\
                    &Solution A(without flare)&Solution B(with flare)&Solution C(with flare)&  \\  
\hline\noalign{\smallskip}
Mode&$~~~~$Semi-detached&$~~~~$Semi-detached&$~~~~$Semi-detached&\\
 $g1=g2$  &$~~~$ 0.32& $~~~~$0.32&   $~~~~$0.32 &  \\
$A1=A2$   &$~~~~$0.5&  $~~~~$0.5&     $~~~~$0.5&    \\
 $  i(deg)        $ &$~~~$ 67.02(13)      & $~~~~$ 66.51(11)   &$~~~~$65.283(10) \\
 
 $q$=$M_{2}/M_{1}$ &  $~~~$ 0.5111(6) & $~~~~$0.5119(3)& $~~~~$0.5113(5)  \\
 $T1$(K)            & $~~~~$3735     &       $~~~~$3735        &     $~~~~$3735          & \\
  $T2$(K)            & $~~~~$3314(21)      &      $~~~~$3216(18)         &     $~~~~$3394(11)          & \\
 $\Omega1$          &$~~~~$0.2897(12)&     $~~~~$0.2898(12)              &          $~~~~$0.2897(19)          & \\
  $\Omega2$          & $~~~~$0.3860(10)&   $~~~~$0.3550(15)                &            $~~~~$0.3760(22)        & \\
  
   $L_{1}/(L_{1}+L_{2})_{B}$  &$~~~~$0.9371(16)&       $~~~~$0.9395(18)    & $~~~~$0.9175(11)  & \\
 $L_{1}/(L_{1}+L_{2})_{V}$  & $~~~~$0.9212(19)&       $~~~~$0.9250(14)     &$~~~~$0.9064(12)& \\
    $L_{1}/(L_{1}+L_{2})_{R}$  &$~~~~$0.9141(17)&      $~~~~$0.9086(15)    &   & \\
 $L_{1}/(L_{1}+L_{2})_{I}$  & $~~~~$0.9021(13)&       $~~~~$0.9043(16)     && \\
 $r1$(pole) & $~~~~$0.4124(1)&$~~~~$0.4123(2)&  $~~~~$0.4124(2)   & \\
 $r1$(side) & $~~~~$0.4378(3)& $~~~~$0.4376(4)&     $~~~~$0.4378(8)          & \\
 $r1$(back) & $~~~~$0.4659(2)& $~~~~$0.4657(2)&    $~~~~$0.4659(9) & \\
 $r2$(pole) & $~~~~$0.1940(4)& $~~~~$0.2197(8)&   $~~~~$0.2280(2) & \\
 $r2$(side) & $~~~~$0.1961(6)& $~~~~$0.2232(1)      &   $~~~~$0.2321(5)           & \\
 $r2$(back) & $~~~~$0.2002(2)&  $~~~~$0.2302(6)            &        $~~~~$0.2404(3)           & \\
 Latitude$_{spot}$(deg)  & $~~~~$20(9)&       $~~~~$47(8)         &  $~~~~$38(8)   & \\
 Longitude$_{spot}$(deg) &$~~~~$316(9)&    $~~~~$247(5)       &  $~~~~$312(9)    & \\
 Radius$_{spot}$(deg)    & $~~~~$17(7)&      $~~~~$25(7)       &   $~~~~$14(7)            & \\
 $T_{spot}/T_{2}$       & $~~~~$0.9589(6)&    $~~~~$0.9588(4)                &   $~~~~$0.980(9)            & \\
\noalign{\smallskip}\hline
  \end{tabular}
  \end{center}
\end{table*}

Then, the  photometric solutions of the light curves of BX Tri with flare activities were calculated based on solution A. 4 nights in total have been observed with flare activities on Nov. 22 and 23, 2014 (named as Solution B with flares) and Nov. 10 and 12, 2017 (named as solution C with flares). In the beginning, we tried to use best-fit results  of solution A to fit the light curves of solution B and C simultaneously, but the results were unsatisfactory. Then, solution B and C were simulated. The initial and basic parameters of Solution B and C are the same as  Solution A, including the effective temperature, the gravity-darkening coefficients, the bolometric albedos and the mass ratio and so on. We still used mode 4 with cool spot of WD program in solution B and C. Finally, the best simulation results were obtained, which clearly shows in Figure 5, open circles and solid lines same as Figure 4.  The detailed photometric solutions of solution B and C are given in Table 5. We can see that the observed data and the theoretical data have a good match in Figure 5. 
Among the simulating progress, the light curves of solution A, B and C must be separately fitted with the cool-spot mode,  meaning that the light curves of BX Tri varies over time. This  phenomenon was also reported by Zhang, Pi \& Yang (2014). After investigating the results of photometric solutions since 2010, we find that almost all of the parameters are similar, except for the position of the cool spots. This implies that BX Tri is an activity system with a long-lived active region on the components.

 

\begin{figure}
 \begin{minipage}[c]{0.5\textwidth}
  \centering
  \includegraphics[width=3.4in]{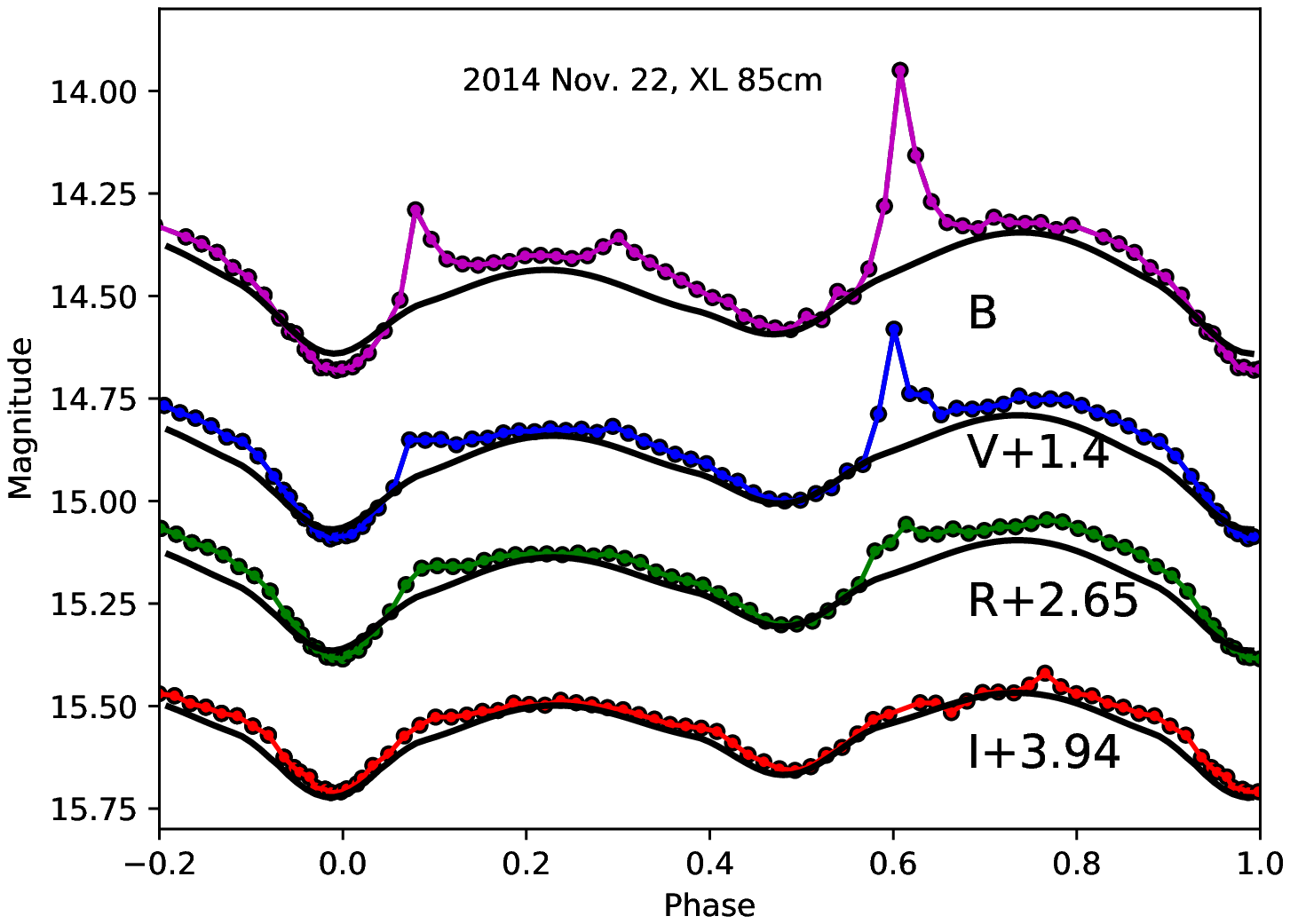}%
  \hspace{1in}%
 \end{minipage}%
 \begin{minipage}[c]{0.5\textwidth}
  \centering
  \includegraphics[width=3.4in]{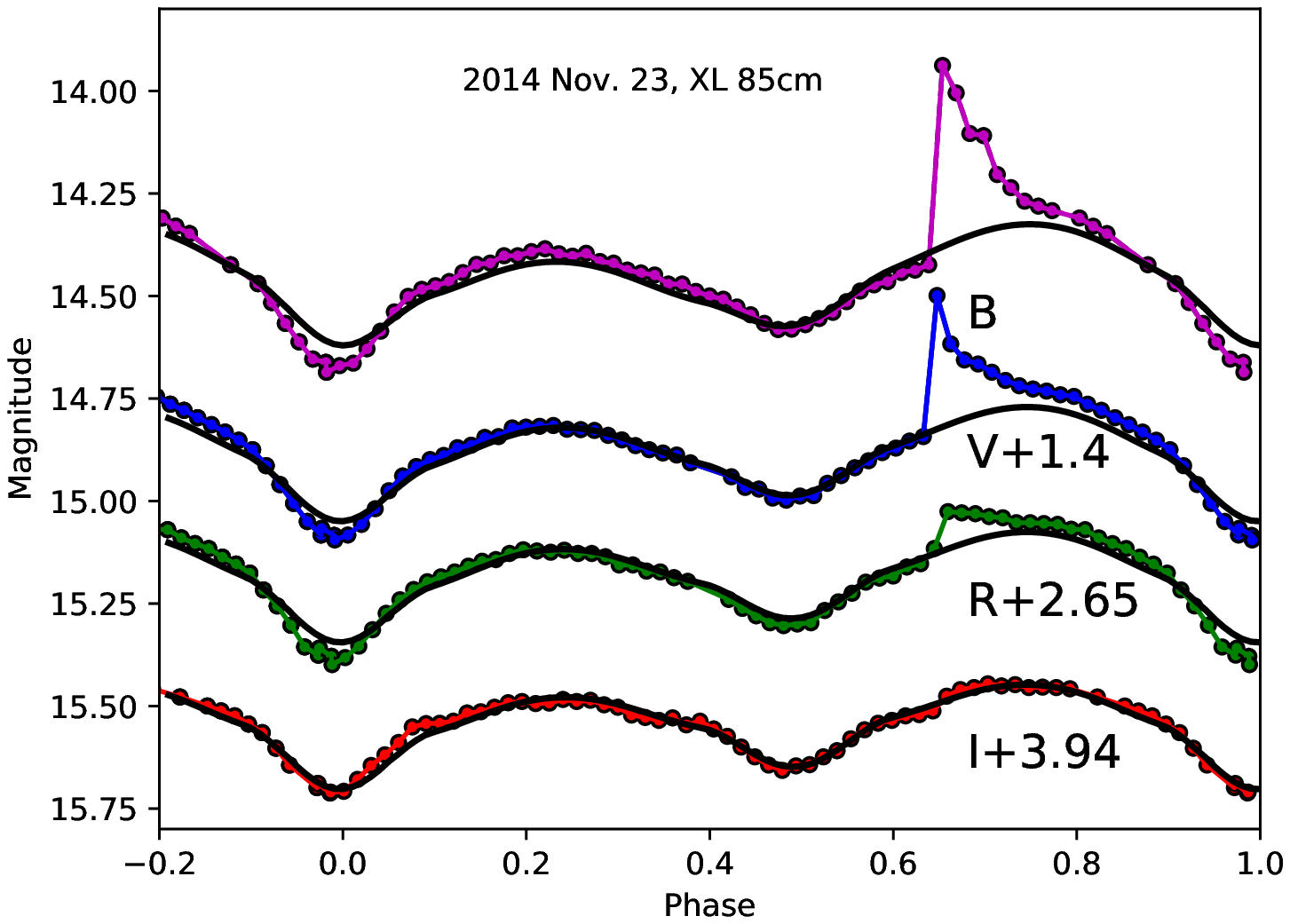}%
  \hspace{0.1in}%
 \end{minipage}%
  \begin{minipage}[c]{0.5\textwidth}
  \centering
  \includegraphics[width=3.4in]{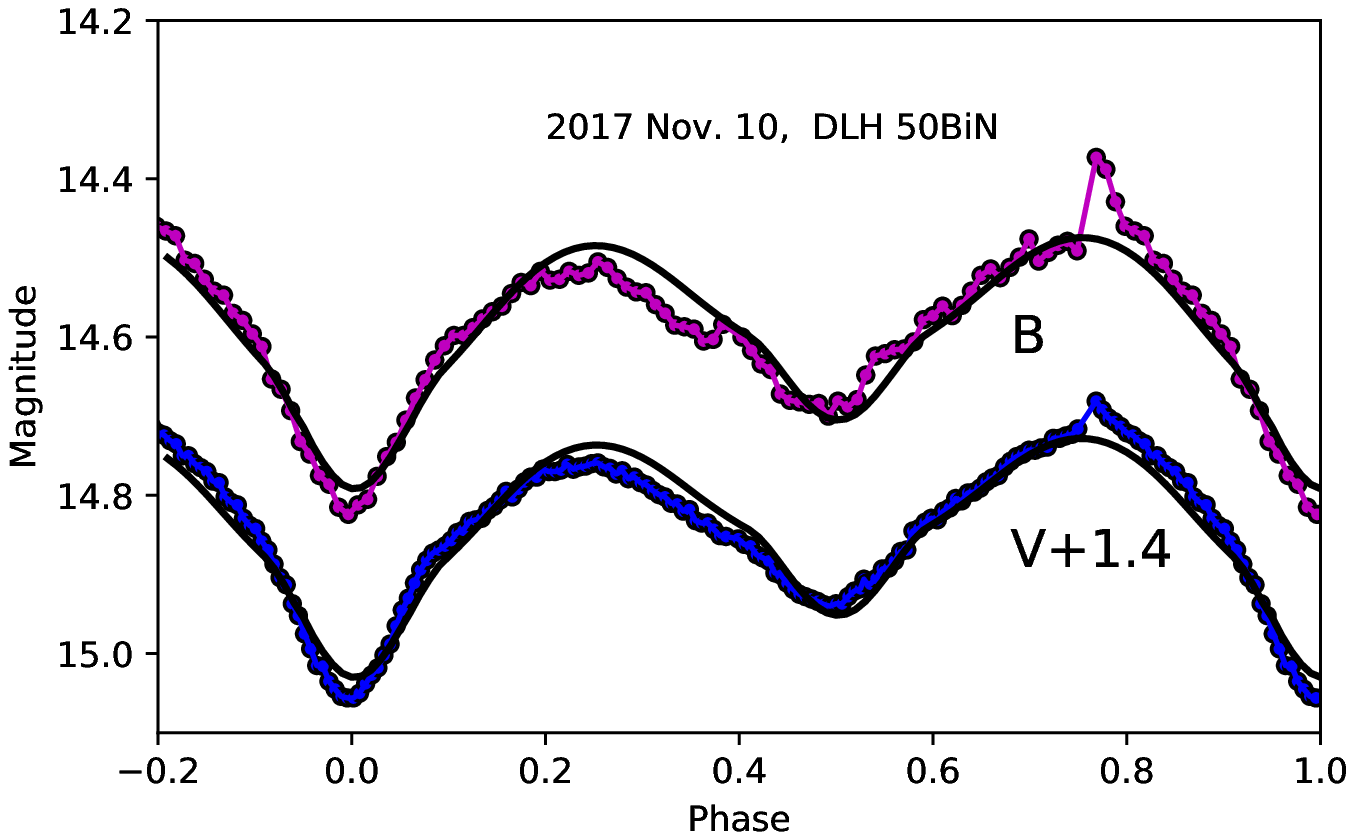}%
  \hspace{1in}%
 \end{minipage}%
 \begin{minipage}[c]{0.51\textwidth}
  \centering
  \includegraphics[width=3.4in]{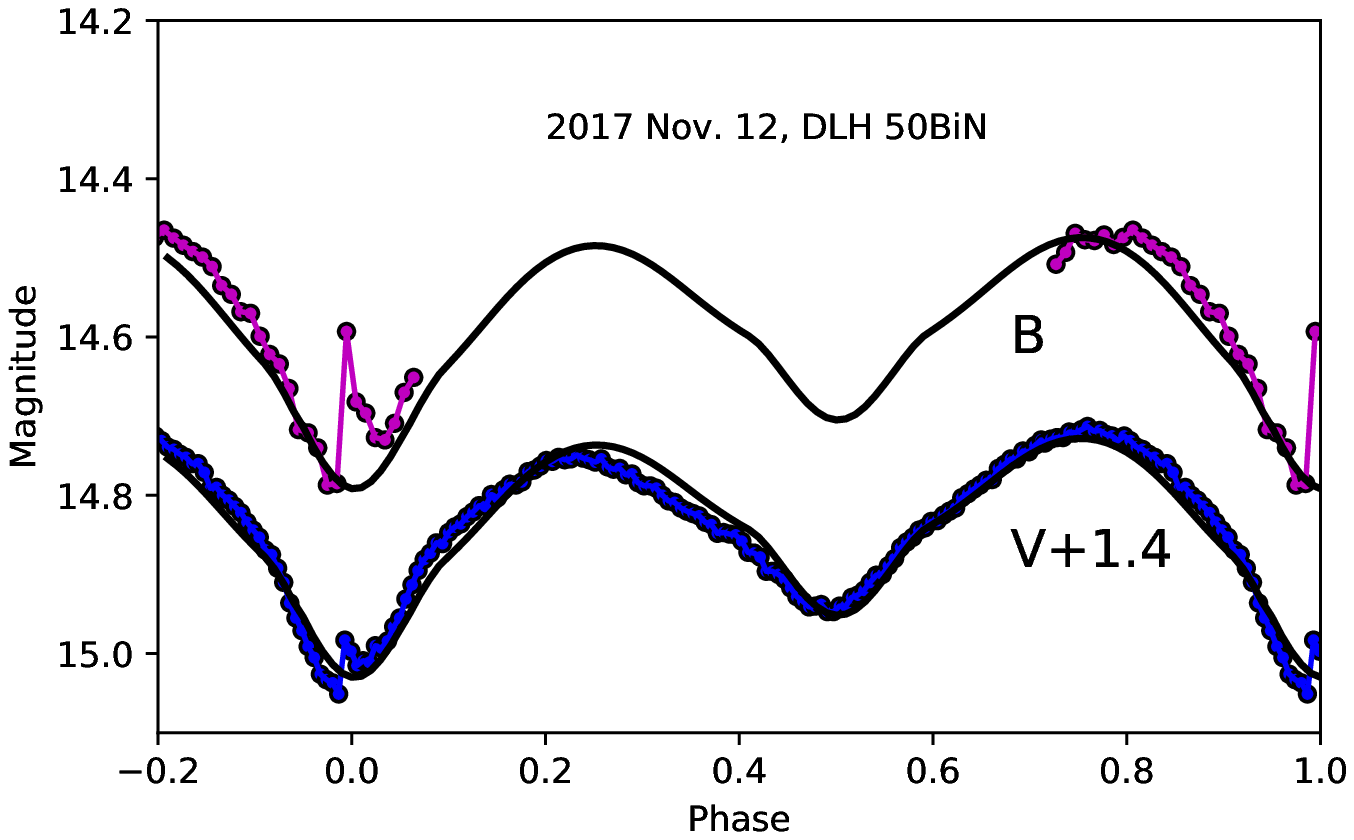}%
  \hspace{0.1in}%
 \end{minipage}%
  \caption{The folded light curves  in B-, V-, R-, and I- bands, loands is displayed with red, green, blue and purple, respectively. The solid line indict the best fit light curve. }
\end{figure}

\section{the flare events}
Flare activity is one typical characteristic of BX Tri. We have collected all of the flare events from literatures and analyzed the statistical properties. 13 flares in total were collected, including 6 flares from our observations and  7 from the literatures (see in Table 6). This is a large sample for an individual object with ground-based multi-color observations. Therefore, we discuss the statistical properties of those flares, including the averaged frequency of flares, flare energies, durations, amplitudes, and the long-term behavior of flare timing with respect to the observed stellar phase.  

\subsection{Flare Frequency}
The average occurrence frequency of flares can be estimated from the number of observed flares and the length of the observation period, which can partly reflect the activity of the star. From recent observations, the best documented dwarf binaries with optical outbursts are GJ 3236 and CM Dra. GJ 3236 was monitored photometrically in all filters for about 900 h with a high flare rate of about 0.06 flares per our. The largest amplitude is 1.3 mag in V band (Smelcer et al. 2017). CM Dra is well known as its high activity. Flare events have been reported by several works (Lacy 1977; Metcalfe, Mathieu \& Latham 1996; Kim et al. 1997; Kozhevnikova et al. 2004). The amplitude range is from 0.02 to 0.7 mag, and the frequency is around 0.02-0.05 flare per hour.

The first observation of a flare in BX Tri was reported by Dimitrov \& Kjurkchieva (2010), who reported 6 flares.  Then, Han et al.(2015) captured another flare event. By the end of Nov, 2017, totally 13 outbursts of this binary have been reported by different observers. The brightness of BX Tri was monitored in total about 144 h between 2010 Nov. to 2017 Nov, including 40 h by Dimitrov \& Kjurkchieva (2010) observations; 24 h by Han et al. (2015) observations and 80 h in this work, respectively.  If we assume that the eruptions are evenly distributed in time, the flare frequency would be about 0.09 flares per hour. This value is larger than those of CM Dra and GJ 3236. An overview of all the observed flares  is given in Table 6. It is shown in our data set that the repeated occurrence of flares lasted for days, and several flare events were detected in a simple night during an orbital cycle. In 2014, the flares lasted two days from Nov. 22  to 23 , and there were 3 events within one period. The period of this system is 0.1926359 days, then we obtained that a flare occurred only about  every 1.54 hours.  In 2017, the first flare was observed on Nov. 10 and the second flare was detected on  Nov. 12.  Although there was no data on Nov. 11 because of the weather, such a flare occurrence indicates that the star has a continue flare outburst.  It suggested that this system has a high flare frequency, which implies strong magnetic activities.

\begin{table*}
  \caption{parameters of observed flares for the BX Tri system. }
   \begin{center}
   \fontsize{7pt}{7pt}
   \selectfont
   \begin{tabular}{llllllllllll}\hline
                 
 Number&   Time &      Filter & phase &  Amplitude     &Duration  &$T_{rise}$$^a$ &$T_{decay}$$^b$ &$L_{peak}$$^c$ &$E_{flare}$$^d$&B-V&Refference \\
              &.                     &           &             & [mag]            &  [sec]       &    [sec]      & [sec]        &[erg/s].     &[erg]         &Mag& \\                
                 
\hline\noalign{\smallskip}
Flare1&2009.10.26 &V      &0.61 &   0.022   & 240         &&&&&&Dimitrov \& Kjurkchieva (2010)\\
Flare2&2009.11.13 &I        &0.64 &   0.085   & 1320        &&&&&&Dimitrov \& Kjurkchieva (2010)\\
Flare3&2009.11.13 &I        &0.84 &   0.027   & 780        &&&&&&Dimitrov \& Kjurkchieva (2010)\\
Flare4&2009.11.13 &R       &0.61 &   0.085   & 1140        &&&&&&Dimitrov \& Kjurkchieva (2010)\\
Flare5&2009.11.13 &R       &0.31 &   0.015   & 540        &&&&&&Dimitrov \& Kjurkchieva (2010)\\
Flare6&2009.11.20 &V       &0.31 &   0.092   & 1500        &&&&&&Dimitrov \& Kjurkchieva (2010)\\
Flare7&2012.12.14 &V     &0.08 &    0.150 &2309       &&&&&&Han et al. 2015 \\

Flare8&2014.11.22& B     & 0.079   & 0.227 &   2268    &566&1702&&&&This paper \\
&& V     & 0.072   & 0.093 &   2552    & 283&2268&&&&This paper \\
&& R     & 0.085   & 0.056 &  2552    &567&1984&&&&This paper  \\
&& I      & 0.101   & 0.035 &   2269    &568&1701&&&&This paper  \\
&&&&&&&&1.79e+31&1.69($\pm$0.11)e+34&0.839&This paper\\

Flare9&2014.11.22& B     & 0.300   & 0.087 &   1985    & 851&1134&&&&This paper \\
&& V     & 0.294   & 0.021 &   1133    & 566&566&&&&This paper \\
&& R     & 0.300   & 0.020 &   568    & 444&123&&&&This paper \\
&& I      & 0.288  & 0.013 &   566    & 284&282&&&&This paper \\
&&&&&&&&4.31e+30&2.38($\pm$0.82)e+33&0.939&This paper\\

Flare10&2014.11.22& B     & 0.607   & 0.476 &   3119    &851&2267& &&&This paper \\
&&V     & 0.600   & 0.283 &   2268    &566&1702&&&& This paper \\
&&R     & 0.614   & 0.096 &   1985    &851&1985&&&& This paper \\
&& I      & 0.629   & 0.015 &   1984    & 1134&850&&&&This paper \\
&&&&&&&&6.38e+31&4.08($\pm$0.24)e+34&0.769&This paper\\

Flare11&2014.11.23& B     & 0.653   & 0.442 &  3232    &498&2733&&&&This paper  \\
&& V     & 0.647  & 0.325 &   3231    &497&2733&&&&This paper  \\
&& R     & 0.644   & 0.013 &   2983    &247&2735&&&&This paper \\
&& I      & 0.673   & 0.015 &   2734    &746&1988&&&&This paper  \\
&&&&&&&&7.77e+31&6.70($\pm$0.31)e+34&0.839&This paper\\

 Flare12&2017.11.10&B     & 0.768   & 0.1035 &   2309    & 495&824&&&&This paper \\     
&&V     & 0.768   & 0.049 &   2413       &629&838&&&&This paper  \\     
&&&&&&&&1.11e+31&8.39($\pm$1.37)e+33&1.091&This paper\\

Flare13&2017.11.12& B& 0.000   & 0.196 &   990         & 330&660&&&&This paper \\
&& V&0.000   & 0.043 &   1049      & 315&734&&&&This paper \\
&&&&&&&&7.58e+30&3.02($\pm$0.84)e+33&1.010&This paper\\

\noalign{\smallskip}\hline
  \end{tabular}
  \end{center}  Notes. a,b),$T_{rise}$ and $T_{decay}$ are the duration of the rise (between the beginning and maximum) and decay (between the maximum and end) phase of the flare, respectively. 
c), $L_{peak}$  is the maximum luminosity of the flare. d), $E_{flare}$ is the total Energy of the flare at the full-wave band..  \end{table*}

\subsection{Flare Duration, Amplitude, Energy and Color}
Flares are usually described by several parameters: amplitude, duration, the equivalent duration and energy. The schematic diagram of a typical flare is showed in Figure 6 based on the flare on 22 Nov. 2014 at B band.  From the shape of the flare in Figure 6, we can see that the flare has a rapid rise and  following by a slowing exponential decay.  For this flare, it just takes about 851 seconds to rise to maximum light and  the decay process  takes 2267 seconds. The duration is 3118 seconds, which is the sum of the rise time and the decay time. The range of the duration for all flares is from 240 sec - 3232 sec. The amplitude is one of the parameters that reflects how strong the flare is. They were calculated by comparing the observed light curve (red line) and the light curve model (black line) with no flares. The light curve model is calculate by using the WD code. It can be calculated as:
 \begin{eqnarray}
\begin{aligned}
Amplitude =M_{peak} - M_{model}, \\ 
\end{aligned}
\end{eqnarray}
where $M_{peak}$ is the flare peak with the lowest magnitude value from observation and $M_{model}$ is the value in the light curve model. In Figure 6, $M_{peak}$ and $M_{model}$ are marked with blue and green circles, receptively. Finally, the amplitude of every flare in different bands were calculated. The difference in different bands is very clearly in Table 6. It clearly shows that the amplitudes of the flares gradually decrease from B to I band. The largest amplitude among those flares is detected in the B band (0.476 mag), V band (0.325 mag), R band (0.056 mag) and I band (0.015mag). The amplitude range of all flares is from 0.013 to 0.476 mag. 

\begin{figure}
\center
\includegraphics[scale=.5]{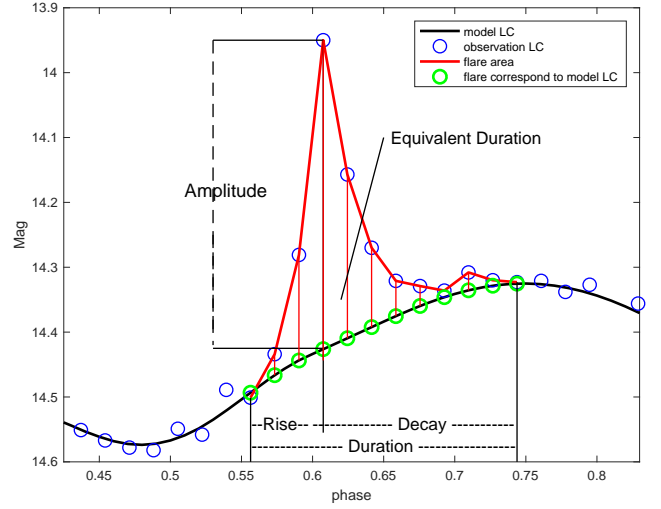}
\caption{One flare light curve in B band observed on Dec 22, 2014 as an example, shows the quantities of amplitude, rise and decay times, and duration.}
\center
\label{fdiff}
\end{figure}


There are many different methods to estimate the energy of a flare. The key is how to get the bolometric luminosity. Recently, many works about flares in Kepler field have been published. They usually assume that the light flare on the star can be  described by a blackbody radiation model (Osten et al. 2016, Kretzschmar 2011, Czesla et al. 2014, Gao et al, 2016) with a specific effective temperature.Then based on the Kepler response function the area of the flare can be obtained. Finally,  the total energy can be calculated by integrating over the duration. 

In this paper, we also use the blackbody radiation model. The difference is that our data are in multiple bands of the Johnson-Cousins system. At first, the basic equation of magnitude was used as follows:
\begin{eqnarray}
\begin{aligned}
m_{f} = -2.5 lg\frac {F_{tot}-F_* }{F_{0}} ~~~~~~~~~~~~~~~~~\\ 
            =-2.5lg(10^{-0.4m_{tot}}-10^{-0.4m_*}),
\end{aligned}
\end{eqnarray}
 where $m$ and F are the visual magnitude in V band and the flux, respectively.  The subscript f, tot, and * represent for the flux from the flare, from both the star and flares, and from the star, respectively. In order to calculate the total energy, we need to know the bolometric luminosity of the flare:
 \begin{eqnarray}
\begin{aligned}
  m_{bol,f} - M_{bol,f} = 5 lg(d) -5,
  \end{aligned}
\end{eqnarray}
where $m_{bol,f}$, $M_{bol,f}$ and d are visual magnitude, absolutely magnitude and distance of the star. $M_{bol,f}$ can be further written as 

 \begin{eqnarray}
\begin{aligned}
  m_{bol,f} = m_{f} +BC_V,
  \end{aligned}
\end{eqnarray}
where $BC_V$ is the bolometric correction in V band. Then we obtained the bolometric luminosity as 
 \begin{eqnarray}
\begin{aligned}
  L _{bol,f}= L_{\odot }\times 10^{0.4(M_{{\odot }}-M_{bol,f})},
  \end{aligned}
\end{eqnarray}
$L_{\odot }$ is the solar luminosity and $M_{{\odot }}$ is the absolute magnitude of the Sun and equal to 4.75 mag. The distance d equals to 95 pc and $BC_V$ equals -1.73 (Dimitrov \& Kjurkchieva ,2010). Combining equations (4)-(7), the bolometric luminosity of the flares $L _{bol,f}$ can be obtained. The peak $L _{bol,f}$  of the flares are presented in Table 6, denoted as $L_{peak}$.  At last, the total bolometric energy of the flares,  $E_{tot,f}$, can be obtained from the integral of $L_{bol,f}$ over the duration:
 \begin{eqnarray}
\begin{aligned}
E_{tot,f} = \int  _{begin} ^{end} L_{ bol,f}(t)  dt.
  \end{aligned}
\end{eqnarray}

The values of the flare energy $E_{tot,f}$ are presented in the tenth column of Table  6. We can see that the range of the total energy is from 3.02($\pm$0.84) $\times$ $10^{33}$ to 6.70($\pm$0.31) $\times$ $10^{34}$ erg, which reach to the superflare level ($10^{34}$ or more, Walkowicz et al. 2011; Hawley et al. 2014; Candelaresi et al. 2014; Chang et al. 2017). \textbf{To obtain the error of the energies,  we therefore perform a Monte Carlo simulation. We first estimate the uncertainties photometry using the deviation from the light curve modeled before and after the flare. Then we assign an additional random Gassion error with sigma of the uncertainty of photometry to each point of flares and estimate anergy with the mock data.  We run this for 100 times and calculate the standard deviation of the energy as the uncertainty.}

Figure 7 shows the  energy - amplitude and  energy - duration correlations with energy error. There are only six points in the figure, because we only calculated the flare energies from our observations. The observed flares (black dots) show a clear power law (solid lines) between energy and amplitude (left panel) and duration (right panel). 

The distributions of rise and decay duration comparing with the total duration are analyzed in all filters (seeing in Figure 8). The red triangles mark the decay durations and the blue dots mark the rise durations. The black lines are linear fits.  They show that the rise duration is nearly constant, while the decay duration increases as the flare duration increases. \textbf{It seemingly means that the flares of BX Tri take almost the same time to reach the peak luminosities, but have different decay times. Of course, this phenomenon need to confirmed by more accurate data, especially on the  time resolution.}

\begin{figure*}
\center
\includegraphics[scale=.6]{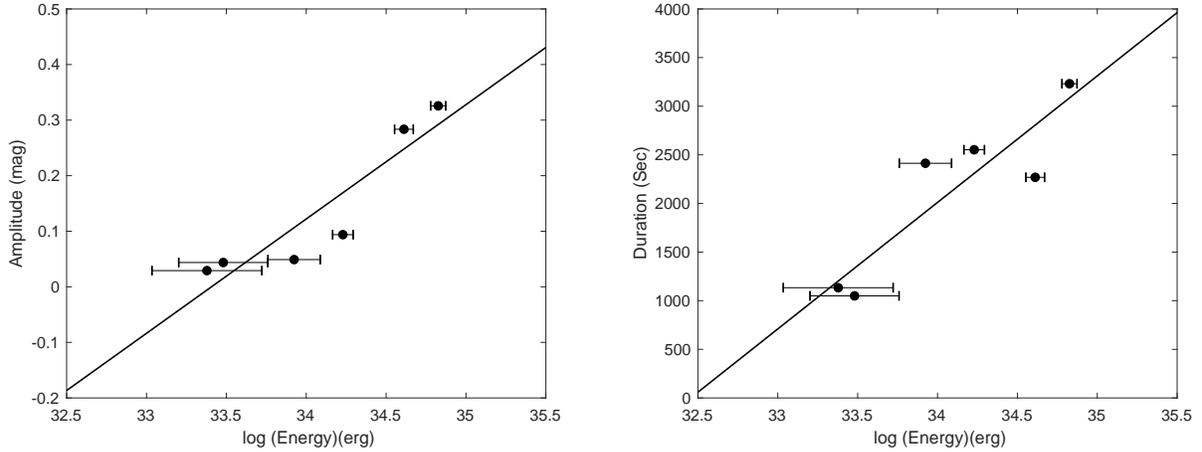}
\center
\caption{Relationships between flare amplitude, energy and duration for BX Tri. The black dots are the observed points, and the solid lines are linear fits.  }
\center
\label{fdiff}
\end{figure*}

\begin{figure}
\center
\includegraphics[scale=.45]{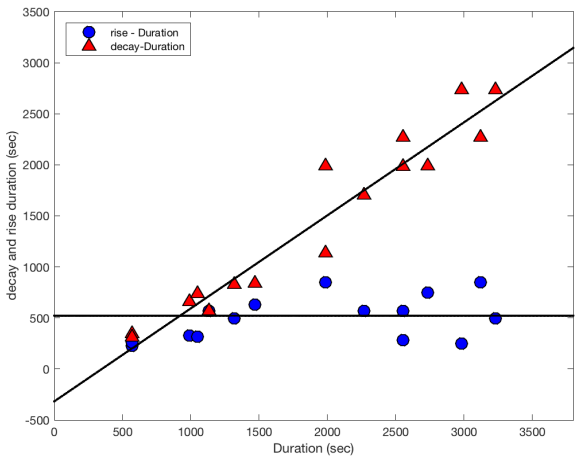}
\center
\caption{The distribution of rise and decay duration comparing with the total duration in all filters. The red triangle marks the decay time and the blue dot makes the rise time. The black line simulate the two duration time. It shows that the rise duration is nearly constant, but the decay duration increases as the flare duration increases. }
\center
\label{fdiff}
\end{figure}

\textbf{Based on the B-band and V-band light curves, the color of  B - V were calculated, shown below the light curve in Figure 1. The change of color during the flare is obvious and the color at the peak of each flare were computed and listed in Table 6. We can see that the B -V  for the two strongest flares are 0.769 mag and 0.839 mag, which is the bluest comparing with the other flares. Tofflemire (2017) were also discussed the color of the flares and showed that the peak emission from a stellar flare is bluer than accretion radiation. They found the peak of the flares is significantly bluer than other measurements that the attributed to accretion. Actually, the bluest color in our paper is even redder than their reddest color. Therefore, the flare of BX Tri system are likely come from accretion. }

 \textbf{On the other hand, we also investigated the distribution of flares as a function of orbital phase, the phase where flares occur are listed in Table 6 and the distribution of these phases is  shown in Figure 9. We can see that the flares occur more at phase $\sim$0.6 and the strongest flare events also happens at this phase. An obvious gap exists at the phase $\sim$0.5,  at which almost all of the light come from the primary. Combining with the color investigate of the flares and period analysis,  all of these information implies that the flares likely occur on the mass  accreting secondary.} 
\begin{figure}
\center
\includegraphics[scale=.45]{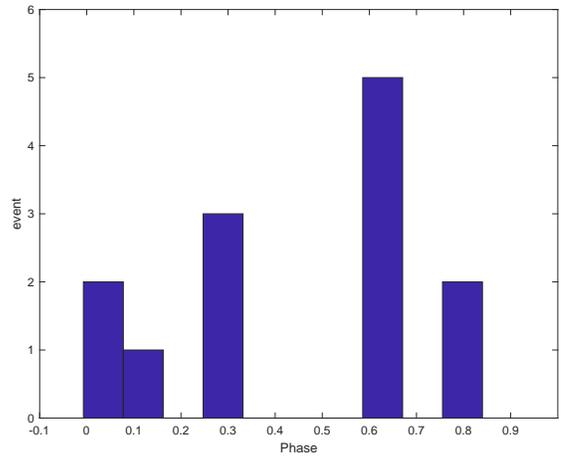}
\caption{Phase distribution of individual flares obtained from all collected flares. We can see that the flare occurred relatively more at phase 0.6 and the strongest flare events also at this phase and have a gap near at phase 0.5.}
\center
\label{fdiff}
\end{figure}

\section{The spectrographic solution of BX Tri}

In addition to photometric monitoring, we also obtained a series of spectroscopic observations with a full phase coverage for BX Tri (see Table 3). The strong emission lines were detected during our observational run, e.g., H$\alpha$, H$\beta$ and H$\gamma$ emission lines, as shown in Figure 2, as well as Ca~{\sc ii} HK lines (but not shown in Figure 2). Such lines (in particular H$\alpha$) are commonly used as chromospheric activity indicators for low-mass stars (e.g., Reid et al. 1995; Bochanski et al. 2007; Walkowicz \& Hawley 2009; West et al. 2015). 
 
As shown in Figure 2, all spectra in different orbital phase show evident emissions in H$\alpha$, H$\beta$ and H$\gamma$ line. The equivalent widths of total emissions of H$\alpha$ and H$\beta$ (EW$_{\text{H}\alpha}$ and EW$_{\text{H}\beta}$ respectively) at each orbital phase were plotted in Figure 10. It shows that the total H$\alpha$ emission varied irregularly in the range of EW$_{\text{H}\alpha}\sim2.5-4$~\AA. On average, EW$_{\text{H}\beta}$ has smaller value, but with similar variation behaviour. It seems that those two Balmer emissions are smaller around the first quadrature (though there are few observations) than those around the second quadrature; A similar H$\alpha$ variation was previously reported by Dimitrov and Kjurkchieva (2010). More interestingly, the H$\beta$ have equivalent width values that comparable to that of H$\alpha$ in several orbital phases (e.g., around the second quadrature), a phenomenon reminiscent of the enhancements in higher Balmer lines during flare-like events (Huenemoerder \& Ramsey 1987; Hawley \& Pettersen 1991; Johns-Krull et al. 1997; Allred et al. 2006; Bochanski et al. 2007), e.g., while H$\beta$ enhanced during a flare-like event, the H$\alpha$ is not proportionally increased in strength. In fact, the ratio of energy emitted in the H$\alpha$ to H$\beta$ is widely used as indicator of the presence of flare-like events; the typical energy ratio value is less than 2 in flare-like events, it then become 3 or larger for quiescent chromospheres (e.g. Huenemoerder \& Ramsey 1987). The EW$_{\text{H}\alpha}$/EW$_{\text{H}\beta}$ with values of 4/4~\AA~around the second quadrature indicates an energy ratio value of about 2, assuming the H$\alpha$ and H$\beta$ continuum flux of BX Tri is comparable to that of a dwarf star with $T_{\text{eff}}=3700$~K. Thus the H$\beta$ enhancements in these phases seem to be due to flare-like events. Unfortunately, we did not have simultaneous photometric observations during these spectroscopic observing runs, here we can not make a solid conclusion that these H$\beta$ enhancements indeed result from flare-like events.

To check the respective contribution to the H$\beta$ enhancements for each component, we fitted each profile of H$\alpha$ and H$\beta$ with a two-Gaussian model, followed the fitting procedure of Dimitrov and Kjurkchieva (2010). We ran a Markov chain Monte Carlo sampling with EMCEE (Foreman-Mackey et al. 2013) to obtain the best-fitted intensities and their uncertainties. The measurements are presented in Figure 10, where red and black symbols denotes the H$\alpha$ (squares) and H$\beta$ (triangles) emissions for the primary and secondary, respectively. Figure 10 shows that the total emissions were mainly from primary component (in particular around the second quadrature) just with one or two exception. In addition, it seems that the variation trend of emission from primary star is opposite from that of emission from secondary, e.g., as the emissions from primary become higher around the second quadrature, the emissions from secondary decrease. For the primary component, there are evident H$\beta$ enhancements in several orbital phases (e.g., 0.661, 0.753), indicating that the primary star might suffer some flare-like events in these phases. For secondary star, there is no evident H$\beta$ enhancements in these phases; instead, we detected evident H$\beta$ enhancements probably due to flare-like events in other phases (e.g., 0.0127). If our measurements for each component are reasonable, it then suggests that flares can occur on both components of BX Tri. However, the flare events, as we have discussed in Section 4, are probably from the secondary star. \textbf{Therefore, more observations are required for the confirmation.}


\begin{figure}
\center
\includegraphics[scale=.45]{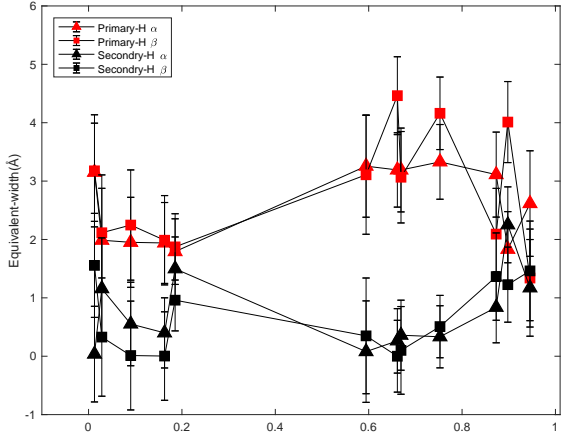}
\caption{Equivalent widths of H$\alpha$, H$\beta$ emission lines of the primary and secondary components at different phases. Red squares and triangles mark H$\alpha$ and H$\beta$ of the primary, respectively. Black squares and triangles mark H$\alpha$ and H$\beta$ of the secondary, respectively.}
\center
\label{fdiff}
\end{figure}

\section{Discussions and Conclusions}

In this paper, new long-term photometric and spectroscopic observations of  the short-period eclipsing system BX Tri have been used to determine its flare activity. Until 2017 December, 13 flares of BX Tri had been reported, 6 of them are from our observations from 2014 December to 2017 December and the other 7 flares are from other published works (Han et al. 2015; Dimitrov et al. 2010). The durations of all flares from all authors ranged from 566 - 3232 sec with amplitudes from 0.013 - 0.476 mag.  If we assume the eruptions of BX Tri are evenly distributed in time, then the frequency is roughly 0.09 flares per hour base on all collected flares. It needs to be emphasized that in our observation on December 22, 2014, 3 flares were detected in one night with one period, which means that a flare occur every 1.54 hours. We can infer that this binary system has a high occurrence frequency of flares comparecd to CM Dra and GJ 3236. 

The total energy of our detected flares have been determined with a range from 3.02($\pm3.02$) $\times 10^{33}$ erg to 6.70($\pm0.31$) $\times 10^{34}$ erg. This energy is up to the superflare energy of $10^{34}$ (e.g., Walkowicz et al. 2011; Notsu et al. 2013; Shibayama et al. 2013; Candelaresi et al. 2014; Hawley et al. 2014; Chang et al. 2017).  The flares on the surface of a star are usually associated with the magnetic energy. The magnetic energy accumulates on the stellar surface accompany with more frequently occurred and larger star spots when the magnetic energy releases flare erupts.  (Shibata $\&$ Magara 2011, Candelaresi et al. 2014). 

Meanwhile, because the photometric solution suggests a semi-detached configuration for BX Tri with the primary star filling the Roche lobe, mass transfer may also affect the flare eruption. Therefore, we analyzed the period variations and  estimated the energy caused by the mass transfer for BX Tri system. The rate of period decreasing of $dP/dt=-1.42 \times 10^{-7}$  $ days~ yr^{-1}$ has been calculated in section 3. The continuous period decreasing can be explained by mass transfer from the primary to secondary.  By considering a conservative mass transfer from primary to secondary, a calculation with the well-known equation,
\begin{eqnarray}
\dot{P}/P = - 3\dot{M_2}(1/M_1-1/M_2)
\end{eqnarray}
leads to a mass transfer rate of $dM_2/dt= 1.31\times 10^{-7} M_\odot~ yr^{-1}$. At last, the luminosity caused by the accreting mass would be approximately represented as the following expressing  (Zhai \& Fang 1995; Zhang et al. 2002):
\begin{eqnarray}
L_{acc} = GM_2 \frac {dm}{dt} (\Omega _2 - \Omega _{in}) /A,
\end{eqnarray}
 where $L_{acc}$ is the accretion luminosity. $M_2$ is the mass of secondary star.  $\frac {dm}{dt}$ means the mass transfer rate. $\Omega _2$ and $\Omega _{in}$ are the dimensionless potentials of the stellar surface and
the critical Roche lobe, A is the distance of the two components, G is the gravitation constant,  respectively.
All of the parameters are known form the photometric results and the period analysis. Finally, the value of $L_{acc}$ was computed as 2.65 $\times 10^{33} erg~s^{-1}$. The average duration of flares can be obtained from our observations as 2113 sec. So the energy caused by the accreting mass can be determined approximately as $E_{acc}$  = 5.6 $\times 10^{36}$ erg .  The largest energy in our observation is $E_{obs}$ = 6.70($\pm0.31$) $\times 10^{34}$ erg. Comparing $E_{acc}$  and $E_{obs}$, we can see that $E_{acc}$ is 100 times bigger than $E_{obs}$, which means the accreting energy form mass transfer is bigger than flare energy from our observation. It can be understand that mass transfer can motivate so large flare outburst but not all of these energy to use to flare outburst. 
 
The energy, amplitude and duration are strongly corrected in Figure 7.  The larger the energy, the greater the amplitude and the longer the duration. If a star has strong magnetic energy, it takes a longer time to release it and then will show a high amplitude. If the magnetic energy is strong enough, a superflare will occur with a higher amplitude and a longer duration. The distributions of rising  and decay time versus the flare duration are analyzed in all filters, showing in Figure 8. The results show that rising time is flat, while the decay time increases with the duration. It implies that the flares of BX Tri 
take almost the same time up to the highest luminosity no matter how long of the duration. Then they decay with the time proportional to the flare duration.

We also investigated the distribution of flares as a function of orbital phase in Figure 9. We can see that the flares occur more frequently at phase 0.6 and the strongest flare events also at this phase.  No flares have been found at phase 0.5, which means all the light is from the primary at this phase.  On the other hand, the results of photometric solutions of BX Tri and the period variation analysis show that this system is a semi-detached binary with a continuous orbital period decrease. BX Tri has a mass transfer from the primary to the secondary. \textbf{Based on these information, we infer that the flares likely occur on the secondary component. Moreover, we identified that the flare attribute to accretion from primary to secondary component based on analysis the B-V color in section 4.2, which seems another evidence to support our point. Nonetheless, more solid and exclusive evidence is still required from good quality, high resolution spectroscopic observation, in particular simultaneous photometric monitoring, in future.}


\textbf{To summaries, the M-type eclipsing binary BX Tri with a short period  is a magnetically active binary system. }By our new  photometric and spectroscopic observations, \textbf{we find that it  is a semi-detached binary with high flare frequency and high energy. In the mean time, the strong emission lines, H$\alpha$, H$\beta$, H$\gamma$ implies that BX Tri has strong chromospheric densities and a strong magnetic activities.}

\section{Acknowledgements} This work is supported by the Beijing Natural Science Foundation 1184018. This work is supported by the National Natural Science Foundation of China (NSFC) and the NSFC/CAS Joint Fund of Astronomy through grants 11803050, 11373037 and U1731111, 11873057. LFM acknowledges the financial support from the  Universidad Nacional Auton\'oma de M\'exico,
under grant PAPIIT IN 100918.

\end{document}